\documentclass[a4paper,fleqn]{cas-dc}

\usepackage{caption}
\usepackage{enumerate}
\usepackage{eufrak}
\usepackage{extarrows}
\usepackage{pifont}
\usepackage{xcolor}

\usepackage{amsthm}
\usepackage{thmtools}
\usepackage{makecell, multirow}
\usepackage{bm}
\usepackage[title]{appendix}
\usepackage{threeparttable}
\usepackage{hyperref}
\usepackage[numbers]{natbib}
\usepackage[subrefformat=parens]{subfig}
\def\tsc#1{\csdef{#1}{\textsc{\lowercase{#1}}\xspace}}
\tsc{WGM}
\tsc{QE}

\newtheorem{theorem}{Theorem}
\newtheorem{definition}{Definition}

\begin{document}
\let\WriteBookmarks\relax
\def\floatpagepagefraction{1}
\def\textpagefraction{.001}

\shorttitle{Blockchain-based Privacy-Preserving Public Key Searchable Encryption with Strong Traceability}
\shortauthors{Y. Han {\em et al.}}

\title [mode = title]{Blockchain-based Privacy-Preserving Public Key Searchable Encryption with Strong Traceability}

\author[a1]{Yue Han}[type=editor,
                        auid=000,bioid=1,
                        style=chinese,
]
\ead{yuehan@seu.edu.cn}
\credit{Methodology, Software, Formal analysis, Writing - Original Draft}
\author[a1]{Jinguang Han}[type=editor,
                        auid=000,bioid=1,
                        style=chinese,
]                        
 \cormark[1] 
\ead{jghan@seu.edu.cn}
\credit{Conceptualization, Writing - Review and Editing, Supervision}
\author[a2]{Weizhi Meng}[type=editor,
                        auid=000,bioid=1,
                        style=chinese,
]
\ead{weme@dtu.dk}
\credit{Investigation, Writing - Review and Editing}
\author[a1]{Jianchang Lai}[type=editor,
                        auid=000,bioid=1,
                        style=chinese,
]
\ead{jclai@seu.edu.cn}
\credit{Visualization, Validation}

\author[a1]{Ge Wu}[type=editor,
                        auid=000,bioid=1,
                        style=chinese,
]
\ead{gewu@seu.edu.cn}
\credit{Validation, Resources}
\address[a1]{School of Cyber Science and Engineering, Southeast University, Nanjing, Jiangsu 210096, China}
\address[a2]{Department of Applied Mathematics and Computer Science, Technical University of Denmark, 2800 Kongens Lyngby, Denmark}

\begin{abstract}
   Public key searchable encryption (PKSE) scheme allows data users to 
    search over encrypted data.
   To identify illegal users, many traceable PKSE schemes have been proposed.
   However, existing schemes cannot trace the keywords which illegal users searched and protect users' privacy simultaneously.
   In some practical applications, tracing both illegal users' identities and the keywords which they searched is quite important to against the
   abuse of data. It is a challenge to bind users' identities and keywords while protecting their privacy.
   Moreover, existing traceable PKSE schemes do not consider the unforgeability and immutability of trapdoor query records,
   which can lead to the occurrence of frame-up and denying.
   In this paper, to solve these problems, we propose a blockchain-based privacy-preserving PKSE with strong traceability (BP3KSEST) scheme. Our scheme provides the following features: (1) authorized users can authenticate to trapdoor generation center and obtain trapdoors without releasing their identities and keywords; (2) when data users misbehave in the system, the trusted third party (TTP) can trace both their identities and the keywords which they searched; (3) trapdoor query records are unforgeable; (4)
   trapdoor query records are immutable because records are stored in blockchain. Notably, this scheme is suitable to the scenarios where privacy must be considered, {\em e.g.}, electronic health record (EHR).
   We formalize both the definition and security model of our BP3KSEST scheme, and present a concrete construction.
   Furthermore, the security of the proposed scheme is formally proven.
   Finally, the implementation and evaluation are conducted to analyze its efficiency.
\end{abstract}




\maketitle
\section{INTRODUCTION}
 \label{sec:sec1}
 Cloud computing delivers computing resource through the Internet and provides cost saving services.
 Nevertheless, data security has become the primary concerns of users.
 To prevent illegal access,  data is encrypted prior to being outsourced to cloud server.
 This brings the search problem over encrypted data.

 Searchable encryption (SE) \cite{Song_2000} is a mechanism which
 enables users to search over encrypted data by using trapdoors for keywords.
 Symmetric searchable encryption (SSE) and public encryption with keyword search (PEKS) are the two main branches of SE.
 In SSE \cite{Song_2000}, only the secret key holder can generate searchable ciphertext and trapdoors.
 In PEKS \cite{Boneh_2004}, 
 keywords are encrypted under the public parameters
 and trapdoors are generated by using the corresponding secret key.

To support flexible access policies,
 PEKS with access control \cite{Jarecki_2009,Xu_2020,liang2015searchable,jiang2016public,cui2016efficient}
 were presented where users need interact with a third party to obtain trapdoors.
Notably, in \cite{Jarecki_2009},
authorized users can obtain trapdoors from trapdoor generation center (TGC) blindly
 without revealing keywords. In \cite{Xu_2020}, a trusted authority (TA) can recover
 keywords after the designated time point, but can not trace data users' identities.

 Existing traceable PEKS schemes \cite{yang2018efficient,yang2021dual,varri2022traceable,zhang2021privacy,Yang_2023}
 cannot trace the identities of illegal user and the keywords which they searched simultaneously.
 However, in some application scenarios, both the identities and keywords need to be traced.
 For example,
 electronic health record (EHR) systems store
 individuals' health related information that are mostly sensitive.
 These personal information must be encrypted before outsourcing
 to cloud servers to reflect the requirements of General Data Protection Regulations (GDPR) \cite{gdpr18}.
 In order to promote data sharing and protect privacy, PEKS
 is combined with EHR
 to enable authorized  users
 to search over  encrypted
 health information. However, when authenticating to the TGC to obtain a trapdoor,
  a  user needs to release his/her identity and the keyword.
 Since both identities and keywords in EHR are sensitive, it will compromise users' privacy. Hence, privacy-preserving PEKS is desirable.
 In a privacy-preserving PEKS scheme, the data user's identity and the keyword in each query should not be revealed.
 Therefore, it is challenging to bind the identities of users and the keywords which they searched.
 Moreover, if users' identities and keywords are not bound, we cannot ensure which keywords a user exactly searched. Therefore, it is necessary to bind users' identities with keywords which they searched.
 Furthermore, schemes \cite{Xu_2020,Yang_2023} did not consider the unforgeability and immutability of trapdoor query records, so  malicious users can erase the trapdoor query records and frame others for searching keywords.
\subsection{Contributions}
 In this paper, we propose a blockchain-based privacy-preserving public key searchable encryption with strong traceability (BP3KSEST) scheme.
 On the one hand, when querying TGC for trapdoors, both users' identities and  keywords
 are protected.
 On the other hand, when data users misbehave in the system,
 a trusted third party (TTP) is able to trace both the identities of malicious users and the keywords which they searched.
 Furthermore, the trapdoor query records are unforgeable and immutable.
 Our scheme provides the following features:
 \begin{itemize}
     \item \textsf{Keyword-Privacy:} the cloud server cannot conclude information about keywords from ciphertext.
     \item \textsf{Keyword-Blindness:} a data user can obtain a trapdoor from TGC without revealing the keyword.
     \item \textsf{Anonymity:} a data user can obtain trapdoors from TGC without releasing anything about his/her identity.
         Moreover, the TGC cannot determine whether
     two trapdoors are generated for the same  user or two different users.
     \item \textsf{Traceability:} only the
    TTP  can de-anonymise data users and trace the keywords queried by them.
     \item \textsf{Unforgeability:} a data user's trapdoor query records cannot be forged by other users, even if they collude with the TTP.
    \item \textsf{Immutability:} data users' trapdoor query records cannot be erased or changed.
 \end{itemize}

 Our main contributions in this paper are summarized as follows:
 (1) a 
 BP3KSEST scheme providing the above features is formally
 constructed;
 (2) both the definition and security model are formalized;
 (3) the security properties of our scheme are formally reduced to well-known complexity assumptions;
 (4) we implement and evaluate our scheme to analyze its efficiency.
 \par

 \subsection{Related work}
 Public-key encryption with keyword search (PEKS) was first proposed by Boneh {\em et al.} \cite{Boneh_2004}.
 That is, all users who know the public parameters can encrypt keywords and send  ciphertext to a cloud server.
 Notably, only the secret key holder can generate trapdoors. After obtaining a keyword ciphertext and a trapdoor,
 the cloud server can test whether they match.
 To support flexible access policy,  PEKS with access
 control \cite{liang2015searchable,jiang2016public,cui2016efficient,Chen_2016,Jarecki_2009,Xu_2020,Yang_2023} were presented.
 In these scheme, to obtain a trapdoor for a keyword, the user must send the keyword to a third party.

 To achieve fine-grained access control and data sharing,
 Liang {\em et al.} \cite{liang2015searchable} proposed a PEKS system
 which supports searchable attribute-based functionality and attribute-based proxy re-encryption.
 Moreover, the system enables a data owner to efficiently share
 data to a specified group of users satisfying a  policy. 
To obtain a trapdoor for a keyword, a data user need interact with the private key generator.

 In scheme \cite{liang2015searchable},
 a single interaction between the data user and the PKG can only obtain one trapdoor.
 In order to reduce the communication cost of generating trapdoors,
 Jiang {\em et al.} \cite{jiang2016public} presented a public key encryption
 with authorized keyword search scheme where data users
 can obtain authorized tokens by sending keyword sets to an authority.
 Then, data users can use the tokens to generate trapdoors for keywords in the  sets.
 Moreover, 
 the size of tokes is constant and independent of the number of keywords.

 To support expressive search predicates, Cui {\em et al.} \cite{cui2016efficient} proposed
 an expressive PEKS scheme which supports conjunctive,
 disjunctive and any monotonic Boolean formulas.
 To obtain a trapdoor, a data user sends
  an access structure over keywords to TGC.
 Then, TGC generates a trapdoor by using his secret key.
 In schemes \cite{liang2015searchable,jiang2016public,cui2016efficient},
 the inside off-line keyword guessing attack (KGA) is an inherent threat.
 \par
 In order to defeat against KGA,
 Chen {\em et al.} \cite{Chen_2016} proposed a server-aided PEKS scheme.
 When generating keyword ciphertexts/trapdoors, the data owner/user needs to interact with the keyword server
 in an authenticated way.
 In this way, the generation of trapdoors and keyword ciphertext must be on-line, hence
 it is secure against inside off-line KGA.
 Moreover, the keyword server can apply rate-limiting measures to resist the on-line KGA.\par

 In PEKS scheme \cite{liang2015searchable,jiang2016public,cui2016efficient,Chen_2016}, data users must reveal keywords to a third party
 to obtain trapdoors, hence the later knows what former search, which compromises data users' keyword privacy. Considering keywords' privacy,
 Camenisch {\em et al.} \cite{Jarecki_2009} proposed a PEKS scheme
  based on blind and anonymous identity-based encryption (IBE) \cite{Boyen_2006},
 where data users can blindly obtain trapdoors from TGC
  without revealing the keywords.

 Based on scheme \cite{Jarecki_2009} and anonymous credential \cite{Au_2006},
 Xu {\em et al.} \cite{Xu_2020} proposed a 
  PEKS with time controlled keyword privacy to analyze query frequency of keywords without destroying the confidentiality
 of  encrypted data and  privacy of users.
 In the scheme \cite{Xu_2020}, there is a time server to generate time tokens.
 When joining the system,  users register to the authorizer and obtain credentials.
 Then authorized  users can obtain trapdoors from TGC
 blindly and anonymously.
 In addition, the
 TA can recover the searched keywords  after the designated time point.
 However, unlike our scheme where both users' identities and keywords can be traced,
 this scheme  only considered the traceability of keywords.
 Moreover, scheme \cite{Xu_2020} did not consider the unforgeability and immutability of trapdoor query records.

In order to trace the users who abuse their secret keys,
Yang {\em et al.} \cite{yang2018efficient} proposed a traceable PEKS scheme, which supports key escrow free and flexible multiple keyword subset search.
In addition, when a data user leaks his secret key,
the data owner can trace the user's identity and revoke his access permission.
However, the scheme \cite{yang2018efficient} cannot trace the identities of data providers.
To solve the problem, Yang {\em et al.} \cite{yang2021dual} proposed a dual traceable distributed attribute-based
searchable encryption.
In \cite{yang2021dual}, a TA can de-anonymise both data providers and data users if necessary.

To prevent the misuse of the secret key, Varri {\em et al.} \cite{varri2022traceable} proposed a traceable and revocable multi-authority PEKS scheme
where the secret keys are generated by two authorities and the TA can trace malicious data users
and revoke them from the system. In scheme \cite{varri2022traceable}, the privacy of the access policy cannot be protected.
To solve this problem,
Zhang {\em et al.} \cite{zhang2021privacy} presented a privacy-preserving attribute-based PEKS with traceability
which supports hidden policy. Moreover, when data users abuse their secret keys, the TA can de-anonymise and revoke them.

 Recently, Yang {\em et al.} \cite{Yang_2023} presented a PEKS scheme, which is time controlled and accountable (traceable)
 where the TA can de-anonymise misbehaved users.
 This scheme supports expressive query policies.
 When joining the system, a  user registers to the TA and obtains a credential with a time 
 period.
  After verifying the credential,
  the time server issues  another credential with a time
   period to the user.
 Subsequently, the user can  blindly and anonymously
 obtain trapdoors from the TGC by using the credentials.
 If users misbehave in the system, the TA can de-anonymise them.

 In recent years, blockchain has been introduced to searchable encryption.
In order to solve the verifier's dilemma in verifiable searchable encryption, Li {\em et al.} \cite{li2021blockchain} presented a blockchain-based searchable encryption with efficient result verification
 and fair payment where the verification is outsourced to the TrueBit network and a fair payment protocol is established
 between multiple data owners and data users. Moreover, data owners can revoke the permission of the document they shared before.

 To enable the sharing of medical data and protect patient privacy,
 Yang {\em et al.} \cite{yang2022blockchain} proposed a
 blockchain-based keyword search scheme with dual authorization for electronic health record sharing, 
 which is key escrow free. Furthermore, an authorization matrix was applied to realize the dual authorization of the identities of users and the corresponding searchable departments.
 Moreover, the MACs of ciphertexts  are stored in blockchain to ensure the integrity of ciphertext.


 Banik {\em et al.} \cite{banik2023blockchain}
 proposed a blockchain-based PEKS scheme which is suitable to share medical data in cloud environment.
 In \cite{banik2023blockchain}, two kinds of blockchains ({\em e.g.} private blockchain and consortium blockchain) were applied.
 The hash of the encrypted electronic health data is stored in the hospital's private blockchain to prevent data from being tampered and
 only the data users who are registered in the consortium blockchain can search.

 To support fine-grained access control and avoid single point of failure,
 Zhang {\em et al.} \cite{zhang2023blockchain} proposed a blockchain-based anonymous attribute-based searchable encryption scheme.
 Because the blockchain have features of tamper-proof, integrity verification and non-repudiation,
 it is responsible for the trusted search results of the search operation. Furthermore, the access policy is hidden
to protect users privacy.

 However, the above blockchain-based searchable encryption scheme \cite{li2021blockchain,yang2022blockchain,banik2023blockchain,zhang2023blockchain} did not
 address the identity traceability and keyword traceability.

 Notably, in schemes \cite{yang2018efficient,yang2021dual,varri2022traceable,zhang2021privacy},
 the TA can only de-anonymise users who abuse their secret keys,
 but cannot trace the identities of users who obtain trapdoors and the searched keywords.
 Schemes \cite{Xu_2020} and \cite{Yang_2023} support the traceability of keywords and  the traceability of users' identities, respectively. However, the unforgeability and immutability of the trapdoor query records were not considered.
 In our scheme, both the identities of users who obtain trapdoors and the keywords which they searched can be traced simultaneously.
 Moreover,  the trapdoor query records in our scheme are unforgeable and immutable.

 In Table \ref{tab:compare}, we compare our scheme with related PEKS schemes in terms of anonymity, blindness,
 identity traceability, keyword traceability, unforgeability, immutability.

 \begin{table*}[!h]
    \caption{The Comparison Between Our Scheme and Related Schemes}
    \label{tab:compare}
    \begin{threeparttable}
    \normalsize
    \resizebox{\textwidth}{!}{
    \begin{tabular}{|c|c|c|c|c|c|c|c|}
    \hline
    Schemes&Anonymity&Blindness&Identity traceability&Keyword traceability&Unforgeability&Immutability\\
    \hline
    \cite{Jarecki_2009}&\ding{55}&\ding{51}&\ding{55}&\ding{55}&\ding{55}&\ding{55}\\
    \hline
    \cite{Xu_2020}&\ding{51}&\ding{51}&\ding{55}&\ding{51}&Credential&\ding{55}\\
    \hline
    \cite{yang2018efficient}&\ding{55}&$\bot$&\ding{51}&\ding{55}&Secret key&\ding{55}\\
    \hline
    \cite{yang2021dual}&\ding{51}&$\bot$&\ding{51}&\ding{55}&Secret key&\ding{55}\\
    \hline
    \cite{varri2022traceable}&\ding{55}&$\bot$&\ding{51}&\ding{55}&Secret key&\ding{55}\\
    \hline
    \cite{zhang2021privacy}&\ding{55}&$\bot$&\ding{51}&\ding{55}&Secret key&\ding{55}\\
    \hline
    \cite{Yang_2023}&\ding{51}&\ding{51}&\ding{51}&\ding{55}&Credential&\ding{55}\\
    \hline
    \cite{li2021blockchain}&\ding{55}&$\bot$&\ding{55}&\ding{55}&\ding{55}&\ding{51}\\
    \hline
    \cite{yang2022blockchain}&\ding{55}&$\bot$&\ding{55}&\ding{55}&\ding{55}&\ding{51}\\
    \hline
    \cite{banik2023blockchain}&\ding{55}&$\bot$&\ding{55}&\ding{55}&\ding{55}&\ding{51}\\
    \hline
    \cite{zhang2023blockchain}&\ding{51}&$\bot$&\ding{55}&\ding{55}&\ding{55}&\ding{51}\\
    \hline
    BP3KSEST&\ding{51}&\ding{51}&\ding{51}&\ding{51}&Trapdoor query record&\ding{51}\\
    \hline
    \end{tabular}}
    \begin{tablenotes}
       \footnotesize
   \item[*] \ding{51}:supported function; \ding{55}:unsupported function; $\bot$:not applicable.
     \end{tablenotes}
   \end{threeparttable}
    \end{table*}

 \subsection{Paper Organization}
 The rest of this paper is organized as follows.
 Section \ref{sec:sec2} presents the preliminaries and notations of
 our scheme. Section \ref{sec:sec3} provides an overview of our scheme.
 Section \ref{sec:sec4} introduces the formal definition and security requirements. The construction of our scheme is given in
 Section \ref{sec:sec5}. Section \ref{sec:sec6} presents the security proof of our scheme. The performance evaluation of our scheme
 is provided in Section \ref{sec:sec7}. Section \ref{sec:sec8} concludes the paper.
 \section{PRELIMINARIES}
 \label{sec:sec2}
 The notation used throughout this paper is summarized in Table \ref{tab:table_notation}.\par

 \begin{table*}[width=0.6\textwidth,pos=!h]
    \caption{Notation Summary}\label{tab:table_notation}
    \centering
    \begin{tabular}{l||l}
        \hline
        Notation & Description \\
        \hline
        $1^l$& Security parameter \\
        $p$ & A prime number\\
        $x \stackrel{R}{\leftarrow} X$ & $x$ is randomly selected from the set $X$\\
        $A(x) \rightarrow y$ & $y$ is computed by running the algorithm $A(\cdot)$ with input x\\
        $\mathcal{BG}(1^l)$ & A bilinear group generation algorithm\\
        $\epsilon(\lambda)$ & A negligible function in $\lambda$\\
        $PPT$ & Probable polynomial-time\\
        $PP$ & Public parameters  \\
        $\mathcal{CA}$ & Center authority \\
        $\mathcal{CS}$ & Cloud server\\
        $\mathcal{DO}$ & Data owner \\
        $\mathcal{DU}$ & Data user \\
        $\mathcal{TGC}$ & Trapdoor generation center\\
        $\mathcal{TR}$ &Tracer\\ 
        $\mathcal{BC}$ &Blockchain\\
        $R_U$ & The trapdoor query record of $\mathcal{DU}$\\
        $ID_U$ & The identity of $\mathcal{DU}$\\
        $\sigma_U$ & The credential of $\mathcal{DU}$\\
        $Table_{\omega}$ & The table that stores keywords\\
        $Table_{ID}$ & The table that stores $\mathcal{DU}$'s identities\\
        \hline
    \end{tabular}
    \end{table*}

 \subsection{Bilinear Group}
     Let $G$ and $G_T$ be cyclic groups with  prime order $p$ and $g$ be a generator of $G$.
     A bilinear pairing map $e:G \times G \to G_T$ satisfies the following properties:
     \begin{enumerate}
    \item{\bf Bilinearity.}  For all $a,b \in G$ and $x,y \in Z_p$, $e(a^x,b^y)=e(a^y,b^x)=e(a,b)^{xy}$.
    \item{\bf No-degeneracy.} For all $a,b \in G$, $e(a,b) \not= 1_T$ where $1_T$ is the identity element in $G_T$.
     \item{\bf Computability.} For all $a,b \in G$, an efficient algorithm exists to compute $e(a,b)$.
     \end{enumerate}
     Let $\mathcal{BG}(1^l) \to (G,G_T,e,p)$ be a bilinear group generator, which takes a security parameter $1^l$ as input and outputs
     a bilinear group $(G,G_T,e,p)$.

 \subsection{Security Assumption}
     \begin{definition} [Decision Diffie-Hellman (DDH) Assumption \cite{DDH1998}]
        Let $G$ be a cyclic group with prime order $p$ and $g$ be a generator of $G$.
     The decision Diffie-Hellman problem is defined as follows.
     Given $(g,g^\alpha,g^\beta,Z\in G)$ for random exponents $\alpha,\beta \in Z_p$, decide whether $Z=g^{\alpha \beta}$
     or $Z=R$ where $R$ is a random element from $G$.
     The DDH assumption holds on $G$
     if all probabilistic polynomial-time (PPT) adversaries $\mathcal{A}$ can solve the DDH problem
     with a negligible advantage,
     namely $Adv^{DDH}_{\mathcal{A}} = |Pr[\mathcal{A}(g,g^\alpha,g^\beta,Z=g^{\alpha \beta})=1]-Pr[\mathcal{A}(g,g^\alpha,g^\beta,Z=R)=1]| \le \epsilon(\lambda)$.\end{definition}

     \begin{definition}[Discrete Logarithm (DL) Assumption \cite{DL1993}]
    Let $G$ be a cyclic group with prime order $p$ and $g$ be a generator of $G$.
     The discrete logarithm problem is defined as follows.
     Given $Z=g^\alpha \in G$ for a random exponent $\alpha \in Z_p$ as inputs, output $\alpha$.
     The DL assumption holds on 
     $G$ if all PPT adversaries $\mathcal{A}$
     can solve the DL problem
     with a negligible advantage, namely $ Adv^{DL}_{\mathcal{A}}=Pr[\mathcal{A}(g,g^\alpha) \to \alpha] \le \epsilon(\lambda)$.
    \end{definition}

     \begin{definition}[LRSW Assumption \cite{lysyanskaya2000pseudonym}]
    Let $\mathcal{BG}(1^l) \to (G,$ $G_T,e,p)$.
     Suppose that $g$ is a generator of $G$.
     The LRSW problem is defined as follows.
     Given $X=g^x, Y=g^y$ for random exponents $x,y \in Z_p$. Suppose that $O_{X,Y}(\cdot)$ is an oracle
     to answer a query on $m \in Z_p$
     with a triple $(a,a^{y},a^{x+mxy})$ for a random group element $a \in G$.
     Next, let $O_{X,Y}(\cdot)$ be called for $m_1,m_2,\ldots m_n$. Then, the problem
     is to generate a quadruple $(m,a,a^{y},a^{x+mxy})$, where $m \in Z_p,m \notin \{0,m_1,m_2,\ldots m_n\} $ and $a \in G$.
     The LRSW assumption
     holds on $(G,G_T,e,p,g)$ if all PPT adversaries $\mathcal{A}$ can solve the LRSW problem with a negligible advantage,
     namely
     $Adv^{LRSW}_{\mathcal{A}} =Pr[\mathcal{A}^{O_{(X,Y)}}(G,G_T,e,p,
     g,g^x,g^y) \to(m,a,b,c):m \notin Q \wedge m \in Z_p \wedge m \neq  0
     \wedge a \in G \wedge b=a^y \wedge c=a^{x+mxy}  ] \le \epsilon(\lambda)$
     , where $Q$ is the set of queries that $\mathcal{A}$
     made to $O_{X,Y}(\cdot)$. \end{definition}

 \subsection{Zero-Knowledge Proof}
     We follow the notation introduced by Camenish and Stadler in \cite{Camenisch_1997} and formalized by Camenish {\em et al.} in \cite{Camenisch_2009}.
     For example, $PoK\{(\mu ,\rho ,\upsilon ):y=g^\mu h^\rho  \wedge \hat{y}=\mathfrak{g}^\rho \mathfrak{h}^\upsilon\}$ denotes a zero-knowledge proof of knowledge
     of integers $\mu,\rho$ and $\upsilon$ such that $y=g^\mu h^\rho$ and $\hat{y}=\mathfrak{g}^\rho \mathfrak{h}^\upsilon$ hold on the
     groups $G=<g>=<h>$ and $\hat{G}=<\mathfrak{g}>=<\mathfrak{h}>$. The convention is that the letters in the parenthesis $(\mu ,\rho ,\upsilon)$
     stand for the knowledge which is being proven, while the other parameters are known by the verifier.

\subsection{Anonymous Credential}
    Anonymous credential is a cryptographic technique that allows a user to prove the possession of a credential without revealing anything more than the
    fact that he owns such a credential. Moreover, the proofs of an anonymous credential is unlinkable.
    The concept of anonymous credential was envisioned by David Chaum \cite{chaum1983blind} and first fully realized by Camenisch and Lysyanskaya \cite{Camenisch_2001}.
    A well-designed anonymous credential should meet the requirements of anonymity, unforgeability and unlinkability.

    A variant of the anonymous credential scheme \cite{Camenisch_2004} is utilized in
    this paper to achieve anonymous trapdoor and search requests. In the following,
    we  review the anonymous credential scheme \cite{Camenisch_2004}.
    \begin{itemize}
      \item $\textsf{CredSetup}(1^l)\to (sk,pk,PP).$ Suppose $(G,G_T,e,p,$ $g)$ is a bilinear group
      where $g$ is a generator of $G$.
      This algorithm selects $x,y\xleftarrow{R} Z_p$ and computes $X=g^x,Y=g^y$.
      The secret key is $sk=(x,y)$, the public key is $pk=(X,Y)$
      and the public parameters are $PP=(G,G_T,e,p,g)$.
      \item $\textsf{CredIssue}(M,\pi_1 ,sk,PP)\to \sigma.$ Given a message $M \in G$ and
       a proof of knowledge $\pi_1:Pok\{(m):M=g^m\}$ where $m \in Z_p$, the credential issuer
       selects $r_u\xleftarrow{R} Z_p$ and computes $a=g^{r_u},b=a^y,c=a^xM^{r_uxy}$.
      The credential on $m$ is $\sigma=(a,b,c)$.

       \item $\textsf{CredProve}(m,\sigma,PP)\to(\tilde{\sigma},\pi_2).$ Given
       a credential $\sigma$,
       a prover selects $r,r'\in Z_p$
       and computes $\tilde{\sigma}=$$(a^{r'},b^{r'},c^{r'r})=$$(\tilde{a},\tilde{b},\tilde{c}^r)$
       $=(\tilde{a},\tilde{b},\hat{c})$. Then, the prover generates a proof of knowledge
       $\pi_2:Pok\{(r,$ $m):v_s^{r^{-1}}=v_xv_{xy}^{m}\}$
       where $v_s=e(g,\hat{c})$,
       $v_x=e(X,\tilde{a})$ and $v_{xy}=e(X,\tilde{b})$.

       \item $\textsf{CredVerify}(\tilde{\sigma},\pi_2,pk) \to 1/0.$ This algorithm outputs 1 if
       $\pi_2$ is valid and $e(\tilde{a},Y)=e(g,\tilde{b})$; otherwise, it outputs 0 to indicate failure.
    \end{itemize}
    The security of this anonymous credential scheme was reduced to the LRSW assumption in the plain model \cite{lysyanskaya2000pseudonym}.

\subsection{TPP-Modulo-Arithmetics}
Two-party protocol for modulo arithmetics (TPP-Modulo-Arithmetics) \cite{Jarecki_2009} is a two-party protocol
for simple arithmetics modulo a prime $p$.
The protocol uses a public key additive homomorphic encryption.
Moreover, the homomorphic encryption should be verifiable,
namely it supports efficient proofs of knowledge
about the encrypted messages.
Camenisch {\em et al.} \cite{Camenisch_2005} presented
an efficient implementation of the protocol by using the Paillier homomorphic
encryption scheme \cite{Paillier_1999}. In \cite{Camenisch_2005},
the  details on how to generate proofs of knowledge are
referred to \cite{camenisch1997proof} and \cite{camenisch2003practical}.
In our scheme, we follow the method proposed in \cite{Jarecki_2009} and \cite{Camenisch_2005}.

\subsection{Blockchain}
Blockchain is a distributed electronic ledger technology, which is based on cryptography, point-to-point network communication technology,
distributed consensus protocol and other technologies \cite{zheng2018blockchain}.
Each block contains a cryptographic hash of the previous block, a timestamp, the current
block's transactions and other information.
Since each block contains the cryptographic hash of the previous block, they form a chain effectively.
Every node on the blockchain runs blockchain consensus protocol and has a copy of ledger.
Once the information of a node is tampered,
its final hash value will be different from the other nodes whose information has not been tampered.
This mechanism guarantees the transparency, immutability and traceability of the blockchain.
In our scheme, trapdoor query records are stored on the blockchain.
This results in the immutability of trapdoor query records.

\section{SCHEME OVERVIEW}
\label{sec:sec3}
\begin{figure*}[!h]
            \centering
            \includegraphics[width=0.95\textwidth]{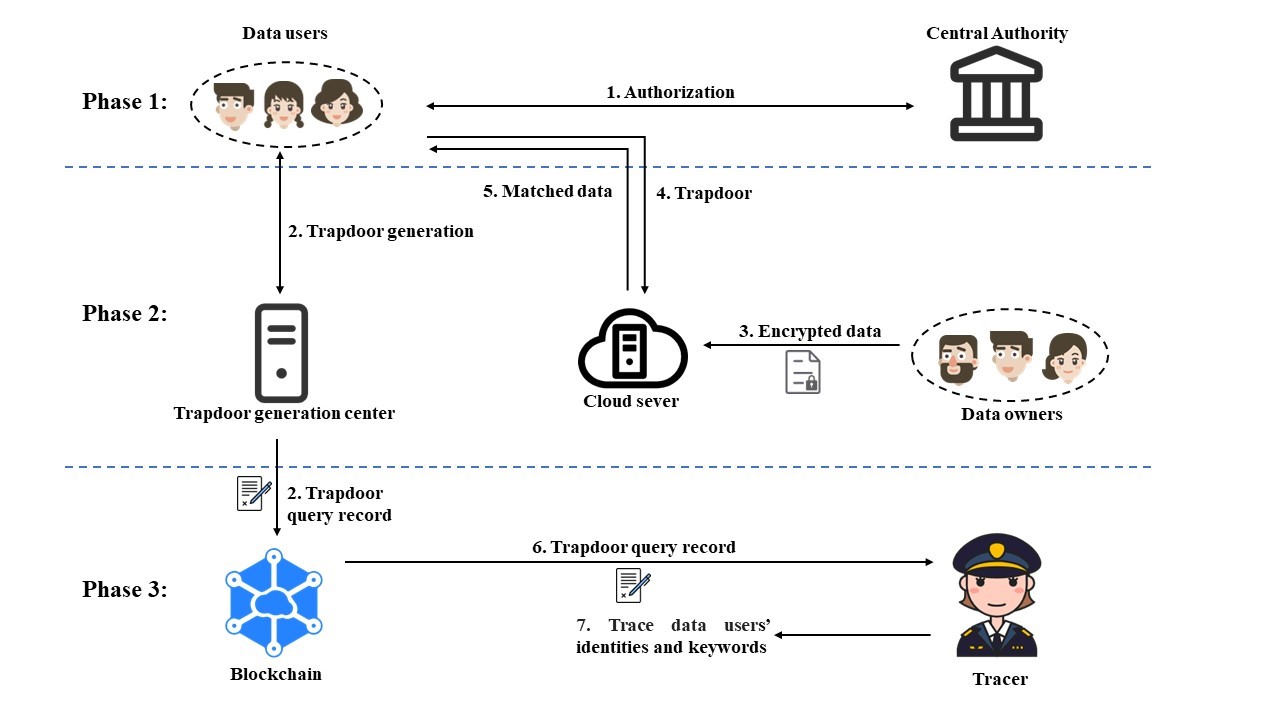}
            \caption{Pictorial description of our scheme}\label{figure:image_1}
        \end{figure*}

Our scheme consists of the following six entities:
\begin{itemize}
    \item a trusted central authority, $\mathcal{CA}$,
     initializes the system and generates a keyword table $Table_{\omega}$.
     Additionally, it issues credentials to data users and
   stores data users' identities $ID_{U}$ and public keys $pk_{U}$ in a table $Table_{ID}$.
   \item a data user, $\mathcal{DU}$, obtains authorization from the central authority and requests trapdoors from the trapdoor generation center.
   Furthermore, the data user can send search requests to the cloud server.
    \item a semi-honest (honest but curious) trapdoor generation center, $\mathcal{TGC}$,
    generates trapdoors for authorized users and stores trapdoor query records $R_{U}$.
    \item a semi-honest (honest but curious) cloud server, $\mathcal{CS}$, 
    performs search operations according to  users' search requests.
    \item data owners, $\mathcal{DO}$,
     uploads encrypted files with encrypted keywords to $\mathcal{CS}$ honestly.
    \item a TTP, $\mathcal{TR}$,
     can de-anonymise $\mathcal{DU}$s and trace keywords which they searched for.
    \item a blockchain, $\mathcal{BC}$, stores the data users' trapdoor query records.
\end{itemize}
A simplified pictorial description of our scheme is presented in Fig.\ref{figure:image_1}.
$\mathcal{CA}$ initializes the system. When joining the system, a $\mathcal{DU}$ authenticates to the $\mathcal{CA}$ and obtains a credential.
To get a trapdoor, $\mathcal{DU}$ sends a trapdoor query record $R_U$ and a proof of the trapdoor request is
generated correctly.
$\mathcal{TGC}$ verifies the proof and generates a blind
trapdoor for $\mathcal{DU}$. $\mathcal{DU}$ can recover a trapdoor from it.
Besides, $\mathcal{TGC}$ stores the trapdoor query record $R_U$ on the blockchain $\mathcal{BC}$.
Then $\mathcal{DU}$ sends the trapdoor to $\mathcal{CS}$.
$\mathcal{DO}$ sends files with encrypted keywords to $\mathcal{CS}$.
By searching,
$\mathcal{CS}$ sends the matched files to $\mathcal{DU}$.
In the case that $\mathcal{DU}$ needs to be traced, given a
trapdoor query record $R_U$ from $\mathcal{BC}$, $\mathcal{TR}$ checks it first.
If the $R_U$ is valid, $\mathcal{TR}$ is able to trace the $\mathcal{DU}$'s identity and the keywords which he searched.\par

        \renewcommand{\figurename}{Fig.}

\section{Formal Definition and Security Requirement}
\label{sec:sec4}
In this section,we review the formal definition and security requirement of our scheme.

\subsection{Formal Definition}
Our scheme is formalized by the following algorithms:
\begin{enumerate}[1)]
    \item $\textsf{Setup}(1^l) \to (PP,Table_{\omega}).$ The algorithm takes a security parameter $1^l$ as input,
    and outputs the public parameters $PP$ and a keyword table $Table_{\omega}$.
    \item $\textsf{KeyGen}.$ This algorithm consists of the following sub-algorithms.
        \begin{itemize}
            \item $\textsf{KeyGen}_\textsf{CA}(PP) \to (sk_{CA},pk_{CA}).$ This algorithm is executed by $\mathcal{CA}$.
           It takes the public parameters $PP$ as inputs and outputs a secret-public key pair $(sk_{CA},pk_{CA})$ for $\mathcal{CA}$.

            \item $\textsf{KeyGen}_\textsf{TGC}(PP) \to (sk_{TGC},pk_{TGC}).$ This algorithm is executed by $\mathcal{TGC}$.
           It  takes the public parameters $PP$ as inputs and outputs a secret-public key pair $(sk_{TGC},pk_{TGC})$ for $\mathcal{TGC}$.

            \item $\textsf{KeyGen}_\textsf{TR}(PP) \to (sk_{TR},pk_{TR}).$ This algorithm is executed by $\mathcal{TR}$.
             It takes the public parameters $PP$ as inputs and outputs a secret-public key pair $(sk_{TR},pk_{TR})$ for $\mathcal{TR}$.

            \item $\textsf{KeyGen}_\textsf{U}(PP) \to (sk_U,pk_U).$ This algorithm is executed by $\mathcal{DU}$s.
            It takes the public parameters $PP$ as inputs and outputs a secret-public key pair $(sk_U,pk_U)$ for $\mathcal{DU}$s.
        \end{itemize}
    \item 
        $\textsf{Reg}(\mathcal{DU}(ID_U,sk_U,pk_U,PP)\leftrightarrow \mathcal{CA}(sk_{CA},PP)) \to (\sigma_U,(ID_U,pk_U)).$ This algorithm is executed between
          $\mathcal{DU}$ and $\mathcal{CA}$. Each $\mathcal{DU}$
           takes his identity $ID_U$,
          secret-public key pair $(sk_U,pk_U)$ and public parameters $PP$ as inputs, and outputs
          a credential $\sigma_U$.
          $\mathcal{CA}$ takes its secret key $sk_{CA}$ and the public parameters $PP$ as inputs, and outputs $(ID_U,pk_U)$. Furthermore, $\mathcal{CA}$ adds $(ID_U,pk_U)$ into a table $Table_{ID}$.
    \item $\textsf{Trapdoor}(\mathcal{DU}(\omega,\sigma_U,sk_U,pk_U,pk_{TR},PP) \leftrightarrow \mathcal{TGC}$ $(sk_{TGC},pk_{CA},PP)) \to (T_\omega,R_U).$ This algorithm is executed between
       $\mathcal{DU}$ and $\mathcal{TGC}$. $\mathcal{DU}$ takes a keyword $\omega$, his credential $\sigma_U$, secret-public key pair $(sk_U,pk_U)$,
          $\mathcal{TR}$'s public key $pk_{TR}$ and the public parameters $PP$ as inputs, and outputs a trapdoor $T_\omega$.
           $\mathcal{TGC}$ takes his secret key $sk_{TGC}$,
          $\mathcal{CA}$'s public key $pk_{CA}$ and the public parameters $PP$ as inputs, and outputs a trapdoor query record $R_U$.
          Then $\mathcal{TGC}$ stores the $R_U$ on $\mathcal{BC}$, which makes trapdoor query records immutable.
    \item $\textsf{PEKS}(pk_{TGC},\omega,PP) \to C.$ This algorithm is executed by $\mathcal{DO}$. $\mathcal{DO}$ takes $\mathcal{TGC}$'s public key $pk_{TGC}$, a keyword $\omega$ and
          public parameters $PP$ as inputs, and outputs the ciphertext $C$ of the keyword $\omega$. 
    \item $\textsf{Test}(T_\omega,C,PP) \to 1/0.$ This algorithm is executed by $\mathcal{CS}$.
    $\mathcal{CS}$ takes a trapdoor $T_\omega$, ciphertext $C$,
          and public parameters $PP$ as inputs, and outputs 1 if the keyword encrypted in $C$ and $T_\omega$ match;
          otherwise, it outputs 0.
    \item $\textsf{Record-Validation}(R_U,pk_{CA},PP) \to 1/0.$ This algorithm is executed by $\mathcal{TR}$.
    $\mathcal{TR}$ takes a trapdoor query record $R_U$,
    $\mathcal{CA}$'s public key $pk_{CA}$ and public parameters $PP$ as inputs,
    and outputs 1 if $R_U$ is valid; otherwise it outputs 0 to show the $R_U$ is invalid.
    \item $\textsf{Trace}(Table_\omega,Table_{ID},R_U,sk_{TR},PP)\to(ID_U,\omega).$ This algorithm is executed by $\mathcal{TR}$.
        It takes the secret key $sk_{TR}$, table $Table_{ID}$ and $Table_\omega$,
    a valid trapdoor query record $R_U$ from $\mathcal{BC}$ and public parameters $PP$ as inputs,
     and outputs an identity $ID_U$ and a keyword $\omega$.
\end{enumerate}

   \begin{definition} A blockchain-based privacy-preserving public key searchable encryption with strong traceability  (BP3KSE\\ST)  is correct if

    \begin{equation*}
    \begin{split}
        \mathsf{Pr}\left[\begin{array}{l|l}
        &\mathsf{Setup}(1^l) \to (PP,Table_\omega);\\
        &\mathsf{KeyGen}_\mathsf{CA}(PP) \to (sk_{CA},\\
        &pk_{CA});\mathsf{KeyGen}_\mathsf{TGC}(PP)\to\\
        &(sk_{TGC},pk_{TGC});\mathsf{KeyGen}_\mathsf{TR}(\\
        &PP)\to(sk_{TR},pk_{TR});\\
        &\mathsf{KeyGen}_\mathsf{U}(PP)\to (sk_U,pk_U);\\
        \mathsf{Test}(T_\omega,C,&\mathsf{Reg}(\mathcal{DU}(ID_U,sk_U,pk_U,PP)\\
        PP)\to 1&\leftrightarrow\mathcal{CA}(sk_{CA},PP))\to(\sigma_U,\\
        &(ID_U,pk_U));\mathsf{Trapdoor}(\mathcal{DU}(\\
        &\omega,\sigma_U,sk_U,pk_U,pk_{TR},PP)\\
        &\leftrightarrow\mathcal{TGC}(sk_{TGC},pk_{CA},PP))\\
        &\to(T_\omega,R_U);\mathsf{PEKS}(pk_{TGC},\\
        &\omega,PP) \to C\\
        \end{array}
        \right]=1
    \end{split}
    \end{equation*}
    and
    \begin{equation*}
    \begin{split}
        \mathsf{Pr} \left[\begin{array}{l|l}
        &\mathsf{Setup}(1^l) \to (PP,Table_\omega);\\
        &\mathsf{KeyGen}_\mathsf{CA}(PP) \to (sk_{CA},\\
        &pk_{CA});\mathsf{KeyGen}_\mathsf{TGC}(PP)\to\\
        &(sk_{TGC},pk_{TGC});\mathsf{KeyGen}_\mathsf{TR}(\\
        &PP)\to(sk_{TR},pk_{TR});\\
        &\mathsf{KeyGen}_\mathsf{U}(PP)\to (sk_U,pk_U);\\
        \mathsf{Trace}(Table_{\omega},&\mathsf{Reg}(\mathcal{DU}(ID_U,sk_U,pk_U,\\
        Table_{ID},R_U&PP)\leftrightarrow\mathcal{CA}(sk_{CA},PP))\to\\
        sk_{TR},PP)&(\sigma_U,(ID_U,pk_U));\mathsf{Trapdoor}(\\
        \to(ID_U,\omega)&\mathcal{DU}(\omega,\sigma_U,sk_U,pk_U,pk_{TR},\\
        &PP)\leftrightarrow\mathcal{TGC}(sk_{TGC},pk_{CA},\\
        &PP))\to(T_\omega,R_U);\mathsf{PEKS}(\\
        &pk_{TGC},\omega,PP) \to C;\\
        &\mathsf{Record-Validation}\\
        &(R_U,pk_{CA},PP) \to 1\\
        \end{array}
        \right]=1.
    \end{split}
    \end{equation*}
\end{definition}

\subsection{Security Requirement}
The security model of our scheme is defined by the following games executed between a challenger $\mathcal{C}$ and an adversary $\mathcal{A}$.\par

$\textbf{Keyword-Privacy.}$ This is used to define the privacy of keywords, namely
$\mathcal{CS}$ cannot conclude information about keywords from ciphertext.
This game is formalized as follows.\par

\textbf{Setup.} $\mathcal{C}$ runs $\textsf{Setup}(1^l) \to (PP,Table_\omega)$ and sends $(PP,Table_\omega)$ to $\mathcal{A}$.\par

\textbf{Phase 1.} $\mathcal{A}$ can adaptively make the following queries.\par

{\bf Key-Generation Query.}
\begin{enumerate}[1)]
    \item $KeyGen_{CA}$ $Query$. $\mathcal{C}$ runs $\textsf{KeyGen}_\textsf{CA}(PP) \to (sk_{CA},$ $pk_{CA})$ and sends $pk_{CA}$ to $\mathcal{A}$.

    \item $KeyGen_{TGC}$ $Query$. $\mathcal{C}$ runs $\textsf{KeyGen}_\textsf{TGC}(PP) \to(sk_{TGC},$ $pk_{TGC})$ and sends $pk_{TGC}$ to $\mathcal{A}$.

    \item $KeyGen_{TR}$ $Query$. $\mathcal{C}$ runs $\textsf{KeyGen}_\textsf{TR}(PP) \to (sk_{TR},$ $pk_{TR})$ and sends $pk_{TR}$ to $\mathcal{A}$.

    \item $KeyGen_U$ $Query$. $\mathcal{C}$ runs $\textsf{KeyGen}_\textsf{U}(PP) \to (sk_U,pk_U)$ and sends $(sk_U,pk_U)$ to $\mathcal{A}$.
\end{enumerate}

{\em Registration Query.} $\mathcal{A}$ adaptively submits a public key $pk_U$ with a zero-knowledge proof and an identity $ID_U$ to $\mathcal{C}$.
$\mathcal{A}$ and $\mathcal{C}$ run $\textsf{Reg}(\mathcal{DU}(ID_U,sk_U,pk_U,PP) \leftrightarrow \mathcal{CA}(sk_{CA},$ $PP)) \to (\sigma_U,(ID_U,pk_U))$. $\mathcal{C}$ returns $\sigma_U$ to $\mathcal{A}$.\par

{\em Trapdoor Query.} $\mathcal{A}$ adaptively submits a trapdoor quest $R_U$ for a keyword $\omega$.
$\mathcal{A}$ and $\mathcal{C}$ run $\textsf{Trapdoor}(\mathcal{DU}(\omega,\sigma_U,$ $sk_U,pk_U,pk_{TR},PP) \leftrightarrow \mathcal{TGC}(sk_{TGC},pk_{CA},PP)) \to (T_\omega,$ $R_U)$.
Then $\mathcal{A}$ obtains $T_{\omega}$. $\mathcal{C}$ stores $R_U$ on the blockchain $\mathcal{BC}$.
Let $T_{trapdoor}$ be an initially empty set which consists of keywords. $\mathcal{C}$ adds $\omega$ into $T_{trapdoor}$.
\par

\textbf{Challenge.} $\mathcal{A}$ submits two keyword $\omega_0$ and $\omega_1$ with restriction of $\omega_0,\omega_1 \notin T_{trapdoor}$.
$\mathcal{C}$ flips an unbiased coin with $\{0,1\}$ and obtains a bit $\mu \in \{0,1\}$.
$\mathcal{C}$ runs $\textsf{PEKS}(pk_{TGC},\omega_{\mu},PP) \to C^{*}$ and sends $C^{*}$ to $\mathcal{A}$.\par

\textbf{Phase 2.} $\mathcal{A}$ continues to make
the key-generation queries,
registration queries and trapdoor queries,
expect the trapdoor queries for $\omega_0,\omega_1$.\par

\textbf{Output.} $\mathcal{A}$ outputs his guess $\mu'$ on $\mu$. $\mathcal{A}$ wins the game if $\mu=\mu'$.\par

\begin{definition} A blockchain-based privacy-preserving public key searchable encryption with strong traceability scheme is $(t,\epsilon(\lambda))$  keyword-privacy if
all PPT adversaries $\mathcal{A}$ can only win the game with a negligible advantage, namely
\begin{align*}
    Adv_{\mathcal{A}}^{keyword-privacy} = \left|Pr[\mu=\mu']-\frac{1}{2} \right| \leq \epsilon(\lambda).
\end{align*}\end{definition}

$\textbf{Keyword-Blindness.}$ This is used to define the blindness of keywords, namely $\mathcal{TGC}$  generates a trapdoor for a keyword without knowing it.
This game is formalized as follows.\par

\textbf{Setup.} $\mathcal{C}$ runs $\textsf{Setup}(1^l) \to (PP,Table_\omega)$ and sends $(PP,Table_\omega)$ to $\mathcal{A}$.\par
\textbf{Key-Generation Query.}
\begin{enumerate}[1)]
    \item $KeyGen_{CA}$ $Query$. $\mathcal{C}$ runs $\textsf{KeyGen}_\textsf{CA}(PP) \to (sk_{CA},$ $pk_{CA})$ and sends $pk_{CA}$ to $\mathcal{A}$.
    \item $KeyGen_{TGC}$ $Query$. $\mathcal{C}$ runs $\textsf{KeyGen}_\textsf{TGC}(PP) \to (sk_{TGC},$ $pk_{TGC})$ and sends $(sk_{TGC},pk_{TGC})$ to $\mathcal{A}$.
    \item $KeyGen_{TR}$ $Query$. $\mathcal{C}$ runs $\textsf{KeyGen}_\textsf{TR}(PP) \to (sk_{TR},$ $pk_{TR})$ and sends $pk_{TR}$ to $\mathcal{A}$.
    \item $KeyGen_U$ $Query$. $\mathcal{C}$ runs $\textsf{KeyGen}_\textsf{U}(PP) \to (sk_U,pk_U)$ and sends $(sk_U,pk_U)$ to $\mathcal{A}$.
\end{enumerate}\par
\textbf{Registration Query.} $\mathcal{A}$ adaptively submits a public key $pk_U$ with a zero-knowledge proof and an identity $ID_U$ to $\mathcal{C}$.
$\mathcal{A}$ and $\mathcal{C}$ run $\textsf{Reg}(\mathcal{DU}(ID_U,sk_U,pk_U,PP)\leftrightarrow \mathcal{CA}(sk_{CA},PP)) \to (\sigma_U,(ID_U,pk_U))$. $\mathcal{C}$ returns $\sigma_U$ to $\mathcal{A}$.\par

\textbf{Trapdoor Query.} $\mathcal{C}$ and $\mathcal{A}$ run $\textsf{Trapdoor}(\mathcal{DU}(\omega,\sigma_U,$ $sk_U,pk_U,pk_{TR},PP) \leftrightarrow \mathcal{TGC}(sk_{TGC},pk_{CA},PP)) \to (T_\omega,$ $R_U)$
where $\mathcal{A}$ runs as $\mathcal{TGC}$ and
obtains $R_U$. Then $\mathcal{A}$ stores $R_U$ on the blockchain $\mathcal{BC}$.  

\textbf{Challenge.} $\mathcal{A}$ submits two keywords $\omega_0$ and $\omega_1$.
$\mathcal{C}$ flips an unbiased coin with $\{0,1\}$ and obtains a bit $\mu \in \{0,1\}$.
$\mathcal{C}$ computes the trapdoor query $R_U$ of $\omega_{\mu}$ and sends $R_U$ to $\mathcal{A}$.
\par

\textbf{Output.} $\mathcal{A}$ outputs his guess $\mu'$ on $\mu$. $\mathcal{A}$ wins the game if $\mu=\mu'$.

\begin{definition}  A blockchain-based privacy-preserving public key searchable encryption with strong traceability is $(t,\epsilon(\lambda))$ keyword-blindness if
all PPT adversaries $\mathcal{A}$ can only win the game with a negligible advantage, namely
\begin{align*}
Adv_{\mathcal{A}}^{keyword-blindness} = \left|Pr[\mu=\mu']-\frac{1}{2} \right| \leq \epsilon(\lambda).
\end{align*} \end{definition}

$\textbf{Unforgeability.}$ This is used to define the unforgeability of data users' trapdoor query record $R_U$,
namely even if $\mathcal{DU}$, $\mathcal{TGC}$ and $\mathcal{TR}$
 collude, they cannot forge a valid trapdoor query record. This game is formalized as follows.\par

\textbf{Setup.} $\mathcal{C}$ runs $\textsf{Setup}(1^l) \to (PP,Table_\omega)$ and sends $(PP,Table_\omega)$ to $\mathcal{A}$.\par
\textbf{Key-Generation Query.}
\begin{enumerate}[1)]
    \item $KeyGen_{CA}$ $Query$. $\mathcal{C}$ runs $\textsf{KeyGen}_\textsf{CA}(PP) \to (sk_{CA},$ $pk_{CA})$ and sends $pk_{CA}$ to $\mathcal{A}$.

    \item $KeyGen_{TGC}$ $Query$. $\mathcal{C}$ runs $\textsf{KeyGen}_\textsf{TGC}(PP) \to (sk_{TGC},$ $pk_{TGC})$ and sends $(sk_{TGC},pk_{TGC})$ to $\mathcal{A}$.

    \item $KeyGen_{TR}$ $Query$. $\mathcal{C}$ runs $\textsf{KeyGen}_\textsf{TR}(PP) \to (sk_{TR},$ $pk_{TR})$ and sends $(sk_{TR},pk_{TR})$ to $\mathcal{A}$.

    \item $KeyGen_U$ $Query$. $\mathcal{C}$ runs $\textsf{KeyGen}_\textsf{U}(PP) \to (sk_U,pk_U)$ and sends $(sk_U,pk_U)$ to $\mathcal{A}$.
\end{enumerate}

\textbf{Registration Query.} $\mathcal{A}$ adaptively submits a public key $pk_U$ with a zero-knowledge proof and an identity $ID_U$ to $\mathcal{C}$.
$\mathcal{A}$ and $\mathcal{C}$ run $\textsf{Reg}(\mathcal{DU}(ID_U,sk_U,pk_U,PP)  \leftrightarrow \mathcal{CA}(sk_{CA},PP)) \to (\sigma_U,(ID_U,pk_U))$.
Let $T_{pk_U}$ be a set consisting of the data users' public key and initially empty.
$\mathcal{C}$ returns $\sigma_U$ and adds $pk_U$ into $T_{pk_U}$. \par

\textbf{Trapdoor Query.} $\mathcal{C}$ and $\mathcal{A}$ run
$\textsf{Trapdoor}(\mathcal{DU}(\omega,\sigma_U,$ $sk_U,pk_U,pk_{TR},PP) \leftrightarrow \mathcal{TGC}(sk_{TGC},pk_{CA},PP)) \to (T_\omega,$ $R_U)$
where $\mathcal{A}$ runs as $\mathcal{TGC}$. Then $\mathcal{A}$ obtains $R_U$ which can be traced to $pk_U$ and stores $R_U$ on the blockchain $\mathcal{BC}$.
$\mathcal{C}$ adds $pk_U$ into $T_{pk_U}$.\par

\textbf{Output.} $\mathcal{A}$ outputs a trapdoor query record $R_U^{*}$ which can be traced to $pk_U^{*}$.
$\mathcal{A}$ wins the game if $\mathsf{Record-Validation}($ $R_U^{*},pk_{CA},PP)=1$
and $pk_U^{*} \notin T_{pk_{U}}$.\par

\begin{definition}  A blockchain-based privacy-preserving public key searchable encryption with strong traceability scheme is $(t,\epsilon(\lambda))$ unforgeable if
all PPT adversaries $\mathcal{A}$ can only win the game with a negligible advantage, namely

\begin{equation*}
\begin{split}
    Adv_{\mathcal{A}}^{unforgeability}&=\text{Pr} \left[\begin{array}{l|l}
        \mathsf{Record-Validation}&pk_U^{*} \notin \\
        (R_U^{*},pk_{CA},PP)=1&T_{pk_{U}}\\
        \end{array}
        \right]\\
        &\leq \epsilon(\lambda).
\end{split}
\end{equation*}

\end{definition}

$\textbf{Traceability.}$ This is used to define the traceability of trapdoor query records,
namely even if a group of $\mathcal{DU}$s, $\mathcal{TGC}$ and $\mathcal{TR}$ collude, they cannot generate a trapdoor query record which cannot be trace
to a member of the colluding group. This game is formalized as follows.\par

\textbf{Setup.} $\mathcal{C}$ runs $\textsf{Setup}(1^l) \to (PP,Table_\omega)$ and sends $(PP,Table_\omega)$ to $\mathcal{A}$.\par

\textbf{Key-Generation Query.}
\begin{enumerate}[1)]
    \item $KeyGen_{CA}$ $Query$. $\mathcal{C}$ runs $\textsf{KeyGen}_\textsf{CA}(PP) \to (sk_{CA},$ $pk_{CA})$ and sends $pk_{CA}$ to $\mathcal{A}$.

    \item $KeyGen_{TGC}$ $Query$. $\mathcal{C}$ runs $\textsf{KeyGen}_\textsf{TGC}(PP) \to (sk_{TGC},$ $pk_{TGC})$ and sends $(sk_{TGC},pk_{TGC})$ to $\mathcal{A}$.

    \item $KeyGen_{TR}$ $Query$. $\mathcal{C}$ runs $\textsf{KeyGen}_\textsf{TR}(PP) \to (sk_{TR},$ $pk_{TR})$ and sends $(sk_{TR},pk_{TR})$ to $\mathcal{A}$.

    \item $KeyGen_U$ $Query$. $\mathcal{C}$ runs $\textsf{KeyGen}_\textsf{U}(PP) \to (sk_U,pk_U)$ and sends $(sk_U,pk_U)$ to $\mathcal{A}$.
\end{enumerate}

\textbf{Registration Query.} $\mathcal{A}$ adaptively submits a public key $pk_U$ with a zero-knowledge proof and an identity $ID_U$ to $\mathcal{C}$.
$\mathcal{A}$ and $\mathcal{C}$ run $\textsf{Reg}(\mathcal{DU}(ID_U,sk_U,pk_U,PP)\leftrightarrow \mathcal{CA}(sk_{CA},PP)) \to (\sigma_U,(ID_U,pk_U))$.
Let $T_{pk_{U1}}$ be a set consisting of the public key selected by $\mathcal{A}$ to make register query and initially empty.
$\mathcal{C}$ returns $\sigma_U$ to $\mathcal{A}$ and adds $pk_U$ into $T_{pk_{U1}}$. \par

\textbf{Trapdoor Query.} $\mathcal{C}$ and $\mathcal{A}$ run $\textsf{Trapdoor}(\mathcal{DU}(\omega,\sigma_U,$ $sk_U,pk_U,pk_{TR},PP) \leftrightarrow \mathcal{TGC}(sk_{TGC},pk_{CA},PP)) \to (T_\omega,$ $R_U)$
where $\mathcal{A}$ runs as $\mathcal{TGC}$.
Then $\mathcal{A}$ obtains $R_U$ which can be traced to $pk_U$ and stores $R_U$ on the blockchain $\mathcal{BC}$.
Let $T_{pk_{U2}}$ be sets consisting of data user's public key selected by $\mathcal{C}$ to answer trapdoor query and initially empty.
$\mathcal{C}$ adds $pk_U$ into $T_{pk_{U2}}$.

\textbf{Output.} $\mathcal{A}$ outputs a trapdoor query record $R_U^{*}$ and $\mathsf{Record-Validation}(R_U^{*},pk_{CA},PP)=1$.
$\mathcal{A}$ wins the game if $\textsf{Trace}(Table_{\omega},$ $Table_{ID},R_U^{*},sk_{TR},PP)\to (ID_U^{*},\omega)$
with $pk_U^{*} \notin T_{pk_{U1}}$ and $pk_U^{*} \notin T_{pk_{U2}}$ or $pk_U^{*} \notin T_{pk_{U1}}$ and $pk_U^{*} \in T_{pk_{U2}}$
where $ID_U^{*}$ and $pk_U^{*}$ are corresponding.\par

\begin{definition}  A blockchain-based privacy-preserving public key searchable encryption with strong traceability scheme  is $(t,\epsilon(\lambda))$ traceable if
all PPT adversaries $\mathcal{A}$ can only win the game with a negligible advantage, namely
\begin{equation*}
    \begin{split}
        Adv_{\mathcal{A}}^{traceability}&=\text{Pr} \left[\begin{array}{c|l}
        &\mathsf{Trace}(\\
        pk_U^{*} \notin T_{pk_{U1}},T_{pk_{U2}}&Table_{\omega},\\
        \mathrm{or}&Table_{ID},\\
        pk_U^{*} \notin T_{pk_{U1}},&R_U^{*},sk_{TR},\\
        pk_U^{*} \in T_{pk_{U2}}&PP)\to\\
        &(ID_U^{*},\omega)\\
        \end{array}
        \right]\\
        &\leq \epsilon(\lambda).
    \end{split}
    \end{equation*}
    where $ID_U^{*}$ and $pk_U^{*}$ are corresponding.
\end{definition}\par

\section{Construction of Our Scheme}
\label{sec:sec5}
\subsection{High-Level Overview}
At a high level, our scheme works as follows.

\textsf{Setup.} $\mathcal{CA}$ initializes system, and generates the public parameters $PP$ and a keyword table $Table_{\omega}$ which includes
$(\omega,g^{\omega})$ for all keywords $\omega$ in the system.

\textsf{KeyGen.} This algorithm has four sub-algorithms.
$\mathcal{CA}$ runs $\textsf{KeyGen}_\textsf{CA}$ and generates secret-public key pairs $(sk_{CA},$ $pk_{CA})$.
$\mathcal{TGC}$ runs $\textsf{KeyGen}_\textsf{TGC}$ and generates secret-public key pairs $(sk_{TGC},pk_{TGC})$.
$\mathcal{TR}$ runs $\textsf{KeyGen}_\textsf{TR}$ and generates secret-public key pairs $(sk_{TR},pk_{TR})$.
Moreover, $\mathcal{DU}$s run $\textsf{KeyGen}_\textsf{U}$ and generate secret-public key pairs $(sk_U,pk_U)$.

\textsf{Registration.} When joining the system, a data user $\mathcal{DU}$ registers with $\mathcal{CA}$ by sending
his identity $ID_U$ and public key $pk_U$.
Meanwhile, the $\mathcal{DU}$ needs make a proof of knowledge $\varPi_1$ to $\mathcal{CA}$
that he knows the secret key $sk_U$ corresponding to $pk_U$.
If $\varPi_1$ is valid, $\mathcal{CA}$ generates the $\mathcal{DU}$'s credential $\sigma_U$
by using secret key $sk_{CA}$.
Finally, $\mathcal{CA}$ sends the credential $\sigma_U$
to $\mathcal{DU}$ and stores $(ID_U,pk_U)$ in $Table_{ID}$.

\textsf{Trapdoor generation.} To get a trapdoor, $\mathcal{DU}$ and $\mathcal{TGC}$
execute a two party protocol TTP-Modulo Arithmetics \cite{Jarecki_2009}.
Then $\mathcal{DU}$ computes $H(\omega)'$, $D_1$, $D_2$, $D_3$, $D_4$, $D_5$ and
randomizes credential $\sigma_U$ to generate anonymous credential $\tilde{\sigma}$.
Note that, $H(\omega)'$ is computed to blind $\omega$.
$(D_1,D_2,D_3)$ are ciphertext of $g^{\omega}$ and $pk_U$.
$(D_4,D_5)$ are used to bind $\mathcal{DU}$'s public key and the keyword which he queried without releasing them.
What's more, data user $\mathcal{DU}$ generates a proof of knowledge $\varPi_2$ to
demonstrate that he is a registered users and the trapdoor request is
generated correctly.
After receiving $H(\omega)'$, $D_1$, $D_2$, $D_3$, $D_4$, $D_5$, $\tilde{\sigma}$ and $\varPi_2$,
$\mathcal{TGC}$ verifies $\varPi_2$ and $\mathcal{DU}$'s anonymous credential $\tilde{\sigma}$.
If they are correct, $\mathcal{TGC}$ computes $(d_0',d_1',d_2',d_3',d_4')$ based on $H(\omega)'$ and sends it to $\mathcal{DU}$.
Finally, $\mathcal{DU}$ can obtain trapdoor $T_{\omega}=(d_0,d_1,d_2,d_3,d_4)$ for $\omega$.
$\mathcal{TGC}$ stores $(H(\omega)', D_1, D_2, D_3, D_4, D_5, \tilde{\sigma},\varPi_2)$ as
a trapdoor query record $R_U$ on the blockchain $\mathcal{BC}$.

\textsf{PEKS.} When encrypting a keyword $\omega$,
 $\mathcal{DO}$ selects $s,s_1,s_2 \xleftarrow{R} Z_p$
and computes the ciphertext $C=(C', C_0, C_1,$ $C_2, C_3, C_4)$.

\textsf{Test.} Given an encrypted keyword $C=(C', C_0, C_1,$ $C_2, C_3, C_4)$ and a trapdoor $T_{\omega}=(d_0,d_1,d_2,d_3,d_4)$,
$\mathcal{CS}$ tests $\prod _{i=0}^{4}e(C_i,d_i) \cdot C' \xlongequal{?} 1$.
If it holds, $\mathcal{CS}$ outputs 1 to indicate the $C$ and $T_{\omega}$ are matched; otherwise, outputs 0 to indicate they are not.

\textsf{Record-Validation.} Given a trapdoor query record $R_U=(H(\omega)', D_1, D_2,
D_3, D_4, D_5, \tilde{\sigma}, \varPi_2)$,
$\mathcal{TR}$ verifies the zero-knowledge proof $\varPi_2$ and credential $\tilde{\sigma}$. If they are valid,
$R_U$ is valid and $\mathcal{TR}$ can trace it; otherwise, $R_U$ is not.

\textsf{Trace.} Given a valid trapdoor query record
$R_U=(H(\omega)', D_1, D_2, D_3, D_4, D_5, \tilde{\sigma},\varPi_2)$ from the blockchain $\mathcal{BC}$,
table $Table_{ID}$ and table
$Table_\omega$, $\mathcal{TR}$ uses his secret key $sk_{TR}$ to decrypt $(D_1,D_3)$ and obtains $g^{\omega}$.
Then $\mathcal{TR}$ looks up $\omega$ corresponding to $g^{\omega}$ in $Table_{\omega}$.
Similarly, $\mathcal{TR}$ uses $sk_{TR}$ to decrypt $(D_2,D_3)$ and obtains $pk_U$.
Finally, $\mathcal{TR}$ looks up $ID_U$ corresponding to $pk_U$ in $Table_{ID}$.

\subsection{Formal Construction}
\renewcommand{\tablename}{Fig.}
The formal construction of our BP3KSEST scheme is formally described
in Figs. \ref{figure:Sfigure2}--\ref{figure:Sfigure9}. \par

\begin{figure*}[!h]
\framebox[\textwidth]{
\parbox{0.95\textwidth}{
\centerline{$\textsf{Setup}(1^l)$}\par
$\mathcal{CA}$ runs $\mathcal{BG}(1^l)\to(G,G_T,e,p)$ with $e:G \times G \to G_T$. Let $g,g_0,g_1,h_1,h_2$ be generators of
$G$. Let $H:x \to g_0g_1^{x}$ be a function from $Z_p$ to $G$. Suppose $H_1:$
$\{0,1\}^{*} \to Z_p$, is a cryptographic
hash function. $\mathcal{CA}$ generates a keyword table $Table_{\omega}$ which includes
$(\omega,g^{\omega})$ for all keywords $\omega$ in this system. $PP=(G,G_T,e,p,g,g_0,g_1,h_1,h_2,H,H_1)$.
}}
\caption{Setup Algorithm}\label{figure:Sfigure2}
\end{figure*}
\begin{figure*}[!h]
\framebox[\textwidth]{
\parbox{0.95\textwidth}{
\centerline{$\textsf{KeyGen}_\textsf{CA}(PP)$}\par
$\mathcal{CA}$ selects $x,y \stackrel{R}{\leftarrow} Z_p$ and computes $X=g^x$, $Y=g^y$. The secret key is $sk_{CA}=(x,y)$ and public key
is $pk_{CA}=(X,Y)$.\par
\centerline{$\textsf{KeyGen}_\textsf{TGC}(PP)$}\par
$\mathcal{TGC}$ selects $t_0,t_1,t_2,t_3,t_4 \stackrel{R}{\leftarrow} Z_p$ and computes
$\Omega=e(g,g)^{t_0 t_1 t_2}$, $\nu_1=g^{t_1}$, $\nu_2=g^{t_2}$, $\nu_3=g^{t_3}$, $\nu_4=g^{t_4}$.
The secret key is $sk_{TGC}=(t_0,t_1,t_2,t_3,t_4)$ and public key is $pk_{TGC}=(\Omega,\nu_1,\nu_2,\nu_3,\nu_4)$.\par
\centerline{$\textsf{KeyGen}_\textsf{TR}(PP)$}\par
$\mathcal{TR}$ selects $x_t\stackrel{R}{\leftarrow} Z_p$ and computes $Y_t=g^{x_t}$. The secret-public key pair is $(x_t,Y_t)$.\par
\centerline{$\textsf{KeyGen}_\textsf{U}(PP)$}\par
$\mathcal{DU}$ selects $x_u\stackrel{R}{\leftarrow} Z_p$ and computes $Y_u=g^{x_u}$. The secret-public key pair is $(x_u,Y_u)$.
}}
\caption{KeyGen Algorithm}\label{figure:Sfigure3}
\end{figure*}

\begin{figure*}[!h]
    \normalsize
    \framebox[\textwidth]{
    \parbox{0.95\textwidth}{
    \normalsize
    \centerline{$\textsf{Reg}(\mathcal{DU}(ID_U,x_u,Y_u,PP)\leftrightarrow \mathcal{CA}(sk_{CA},PP))$}\par
    \begin{tabular}{lcl}
    Data User: $\mathcal{DU}$& &\ Central Authority: $\mathcal{CA}$\\
    Compute the proof & &\ Verify $\varPi _1$.\\
    $\varPi _1:PoK\{(x_u):Y_u=g^{x_u}\}$&$\xrightarrow{ID_U,Y_u,\varPi _1}$&\ Select $r_u\xleftarrow{R}Z_p$, and compute $a=g^{r_u}$, $b=a^y$, $c=a^{x}Y_u^{r_uxy}$.\\
    Keep $\sigma_U$ as credential.&$\xleftarrow{\sigma_U}$&\ Let $\sigma_U=(a,b,c)$ and store $(ID_U,Y_u)$ in $Table_{ID}$.\\
    \end{tabular}}}
    \caption{Registration Algorithm}\label{figure:Sfigure4}
    \end{figure*}

\begin{figure*}[!h]
\framebox[\textwidth]{
\parbox{0.95\textwidth}{
\centerline{$\textsf{Trapdoor}(\mathcal{DU}(\omega,\sigma_U,x_u,Y_u,Y_t,PP) \leftrightarrow \mathcal{TGC}(sk_{TGC},pk_{CA},PP))$}\par
\begin{tabular}{lcl}
\normalsize
Data User: $\mathcal{DU}$& &Trapdoor generation center: $\mathcal{TGC}$\\
Select $u_0,u_1,u_2,r_1',r_2' \xleftarrow{R} Z_p$. & & Select $\hat{r_1},\hat{r_2}\xleftarrow{R} Z_p$. \\
&$\xleftrightarrow[Arithmetics]{TTP-Modulo}$ &  Outputs: \\
&&$x_0=(\hat{r_1}r_1't_1t_2+\hat{r_2}r_2't_3t_4)+u_0$;\\
&&$x_1=-(u_3/r_1' \cdot t_0 t_2 )+u_1$;\\
&&$x_2=-(u_3/r_1' \cdot t_0t_1)+u_2$;\\
Select $r_0,r,r',u_3 \xleftarrow{R} Z_p$.&&\\
Compute $H(\omega)' =(g_0g_1^{\omega})^{u_3}$, $D_1=g^{\omega}Y^{r_0}_t$, $D_2= Y_uY^{r_0}_t$,&&\\
$D_3=g^{r_0}$, $D_4=g^{\omega}h^{r}_1$, $D_5=Y_uh^{r}_2$, $\tilde{\sigma}=(a^{r'},b^{r'}, c^{r'r})=$&&\\
$(\tilde{a},\tilde{b},\hat{c})$. &&\\
Compute the proof $\varPi _2:$&&\\
$PoK\{(\omega,u_3,r_0,r,x_u):H(\omega)' =(g_0g_1^{\omega})^{u_3}\wedge D_1=g^{\omega}Y^{r_0}_t\wedge $&&\\
$D_2= g^{x_u}Y^{r_0}_t \wedge D_3=g^{r_0}\wedge D_4=g^{\omega}h^{r}_1 \wedge D_5=g^{x_u}h^{r}_2\wedge$&&\\
$v^{r^{-1}}_s=v_x v^{x_u}_{xy}\}$ where $v_s=e(g,\hat{c})$, $v_x=e(X,\tilde{a})$&&\\
and $v_{xy}=e(X,\tilde{b})$. &&\\
&$\xrightarrow[D_4,D_5,\tilde{\sigma},\varPi _2]{H(\omega)',D_1,D_2,D_3}$ & Verify $\varPi_2$ and $e(\tilde{a},Y)\xlongequal{?}e(g,\tilde{b})$.\\
&&Compute $d_0'=g^{x_0}$, $d_1'=g^{x_1}H(\omega)'^{-\hat{r_1}t_2}$,\\
&&$d_2'=g^{x_2}H(\omega)'^{-\hat{r_1}t_1}$, $d_3'=H(\omega)'^{-\hat{r_2}t_4}$, \\
&&$d_4'=H(\omega)'^{-\hat{r_2}t_3}$.\\
&&Store $R_U=(H(\omega)',D_1,D_2,D_3,D_4,D_5,\tilde{\sigma},\varPi _2)$\\
&&on $\mathcal{BC}$.\\

Compute $d_0=d_0'\cdot g^{-u_0}$, $d_1=(d_1'\cdot g^{-u_1})^{\frac{r_1'}{u_3}}$,&$\xleftarrow{(d_0',d_1',d_2',d_3',d_4')}$&\\
$d_2=(d_2' \cdot g^{-u_2})^{\frac{r_1'}{u_3}}$, $d_3=(d_3')^{\frac{r_2'}{u_3}}$, $d_4=(d_4')^{\frac{r_2'}{u_3}}$.&&\\
Keep $T_{\omega}=(d_0,d_1,d_2,d_3,d_4)$.&&\\
\end{tabular}
}}
\caption{Trapdoor Generation Algorithm}\label{figure:Sfigure5}
\end{figure*}

\begin{figure*}[!h]
\framebox[\textwidth]{
\parbox{0.95\textwidth}{
\centerline{$\textsf{PEKS}(pk_{TGC},\omega,PP)$}\par
Given a keyword $\omega$, $\mathcal{DO}$ works as follows:
(1)Select $s,s_1,s_2 \xleftarrow{R} Z_p$. (2)Compute  $C'=\Omega^s$, $C_0=H(\omega)^s$, $C_1=\nu_1^{s-s_1}$, $C_2=\nu_2^{s_1}$, $C_3=\nu_3^{s-s_2}$,$C_4=\nu_4^{s_2}$.
(3)Output $C=(C',C_0,C_1,C_2,C_3,C_4)$.
}}
\caption{PEKS Algorithm}\label{figure:Sfigure6}
\end{figure*}

\begin{figure*}[!h]
\framebox[\textwidth]{
\parbox{0.95\textwidth}{
\centerline{$\textsf{Test}(T_{\omega},C,PP)$}\par
Given a trapdoor $T_{\omega}=(d_0,d_1,d_2,d_3,d_4)$ and a encrypted keyword $C=(C',C_0,C_1,C_2,C_3,C_4)$,
$\mathcal{CS}$ checks $\prod_{i=0}^{4} e(C_i,d_i) \cdot C' \xlongequal{?} 1$.
If it holds, output $1$; otherwise, output $0$.
}}
\caption{Test Algorithm}\label{figure:Sfigure7}
\end{figure*}
\begin{figure*}[!h]
\framebox[\textwidth]{
\parbox{0.95\textwidth}{
\centerline{$\textsf{Record-Validation}(R_U,pk_{CA},PP)$}\par
Given a trapdoor query record $R_U=(H(\omega)',D_1,D_2,D_3,D_4,D_5,\tilde{\sigma},\varPi_2)$,
$\mathcal{TR}$ verifies zero-knowledge proof $\varPi_2$ and $e(\tilde{a},Y) \xlongequal{?} e(g,\tilde{b})$.
If they are valid, the algorithm outputs 1 to indicate the $R_U$ is valid;
otherwise, it outputs 0 to indicate the $R_U$ is not.
}}
\caption{Record Validation Algorithm}\label{figure:Sfigure8}
\end{figure*}
\begin{figure*}[!h]
\framebox[\textwidth]{
\parbox{0.95\textwidth}{
\centerline{$\textsf{Trace}(Table_{\omega},Table_{ID},R_U,x_t,PP)$}\par
Given a valid trapdoor query record $R_U=(H(\omega)',D_1,D_2,D_3,D_4,D_5,\tilde{\sigma},\varPi _2)$ from $\mathcal{BC}$, $\mathcal{TR}$ works as follows:
(1) Compute $g^\omega=\frac{D_1}{D_3^{x_t}}$ and $Y_u=\frac{D_2}{D_3^{x_t}}$. (2) Look up $Table_{\omega}$ and $Table_{ID}$.
(3) Trace $\omega$ and $ID_U$ corresponding to $g^{\omega}$ and $Y_u$, respectively.
}}
\caption{Trace Algorithm}\label{figure:Sfigure9}
\end{figure*}

\subsection{Correctness}
The details of the zero-knowledge proofs of $\varPi_1$ and $\varPi_2$ are provided in Appendix
\ref{section:append1} and Appendix \ref{section:append2}.\par
Our scheme is correct as the following equations hold.
In the trapdoor generation and record valid algorithm, the credentials can be verified by the following equations.\par
\begin{align*}
&e(\tilde{a},Y)= e(g^{r_ur'},g^y)=e(g,g^{r_ur'y})=e(g,a^{yr'})\\
&=e(g,\tilde{b})
\end{align*}
and
\begin{align*}
&e(X,\tilde{a}) \cdot e(X,\tilde{b})^{x_u} =e(g^x,a^{r'}) \cdot e(g^x,b^{r'x_u})\\
&=e(g,a^{xr'}) \cdot e(g,b^{x_u  x r'})=e(g,a^{xr'}) \cdot e(g,a^{y  x_u  x r'})\\
&=e(g,a^{xr'+ x_u x y r'})=e(g,\hat{c})^{r^{-1}}.
\end{align*}\par
In the trapdoor generation and test algorithm, the correctness of keyword search can be verified by the following equations.\par
\begin{align*}
&\prod^{4}_{i=0} (C_i,d_i)\cdot C' =e(H(\omega)^s, g^{\hat{r_1} r'_1 t_1 t_2 + \hat{r_2} r'_2 t_3 t_4})\cdot\\
&e(\nu^{s-s_1}_1,(g^{x_1-u_1}H(\omega)^{'-\hat{r_1}t_2})^\frac{r'_1}{u_3}) \cdot\\
&e(\nu^{s_1}_2,(g^{x_2-u_2}H(\omega)^{'-\hat{r_1}t_1})^{\frac{r'_1}{u_3}})\cdot \\
&e(\nu^{s-s_2}_3,H(\omega)^{' \frac{-\hat{r_2} r'_2 t_4}{u_3}}) \cdot e(\nu^{s_2}_4, H(\omega)^{' \frac{-\hat{r_2} r'_2 t_3}{u_3}}) \cdot \varOmega  ^ {s} \\
={}&e(H(\omega)^{s},g^{\hat{r_1} r'_1 t_1 t_2 + \hat{r_2} r'_2 t_3 t_4})\cdot\\
&e(g^{t_1(s-s_1)},g^{-t_0 t_2}H(\omega)^{-\hat{r_1} r'_1 t_2 }) \cdot \\
&e(g^{t_2s_1},g^{-t_0 t_1} H(\omega)^{-\hat{r_1} r'_1 t_1})\cdot \\
&e(g^{t_3(s-s_2)},H(\omega)^{-\hat{r_2} r'_2 t_4}) \cdot e(g^{t_4s_2}, H(\omega)^{{-\hat{r_2} r'_2 t_3}})\cdot\\
&e(g,g)^{t_0 t_1t_2 s}\\
={}&e(g,H(\omega))^{s\cdot(\hat{r_1} r'_1 t_1 t_2 + \hat{r_2} r'_2 t_3 t_4)}\cdot e(g,g)^{-t_0 t_1 t_2 (s-s_1)} \cdot\\
&e(g,H(\omega))^{-\hat{r_1} r'_1 t_1 t_2(s-s_1)}\cdot e(g,g)^{-t_0 t_1 t_2 s_1} \cdot\\
&e(g,H(\omega))^{-\hat{r_1} r'_1 t_1 t_2 s_1}\cdot e(g,H(\omega))^{-\hat{r_2} r'_2 t_3 t_4 (s-s_2)} \cdot\\
&e(g,H(\omega))^{-\hat{r_2} r'_2 t_3 t_4 s_2}
\cdot e(g,g)^{t_0t_1 t_2 s}\\
={}&e(g,g)^{0} \cdot e(g,H(\omega))^{0}\\
={}& 1.
\end{align*}

\section{Security Proof}
\label{sec:sec6}

In this section, the security of our scheme is formally proven.\par

\begin{theorem} Our scheme is $(t',\epsilon'(\lambda))$-keyword-privacy if the IBE scheme \cite{Boyen_2006} is $(t,\epsilon(\lambda))$-IBE-ANO-CPA security where $\epsilon(\lambda)\geq \epsilon'(\lambda)$ and $t=O(t')$.
\end{theorem}\par

\textit{Proof:} The PEKS of our scheme is constructed based on the anonymous IBE scheme \cite{Boyen_2006} which has been proved to be IBE-ANO-CPA security.
We follow the transformation ibe-2-peks \cite{Boneh_2004} which takes
as input a secure IBE scheme and returns a PEKS scheme.
The IBE scheme \cite{Boyen_2006} consists of four algorithms:
$\textsf{Setup}_\textsf{I}(1^l)\to(mpk_{I},msk_{I},PP_{I})$, $\textsf{Extract}_\textsf{I}(msk_{I},id,PP_{I}) \to sk_{id}$,
$\textsf{Encrypt}_\textsf{I}(mpk_{I},id,M,$ $PP_{I})\to C$, $\textsf{Decrypt}_\textsf{I}(C,sk_{id},PP_{I}) \to M$.
In our scheme, the $sk_{TGC}$ and $pk_{TGC}$ are $msk_{I}$ and $mpk_{I}$, respectively.
In our $\textsf{PEKS}(pk_{TGC},\omega,PP) \to C$ algorithm, $C$ is the ciphertext of
the identity element 1 in $G_T$, namely $\textsf{Encrypt}_\textsf{I}(pk_{TGC},$ $\omega,1,PP)=\textsf{PEKS}(pk_{TGC},\omega,$ $PP)$.
The trapdoor $T_{\omega}$ of a keyword $\omega$ in our PEKS scheme is the secret key corresponding to identity
$\omega$ in IBE scheme \cite{Boyen_2006}, namely $\textsf{Extract}_\textsf{I}(sk_{TGC},$ $\omega,PP)\to T_{\omega}$.
$\textsf{Test}(T_\omega,C)$ in our scheme is implemented by executing verifying $\textsf{Decrypt}_\textsf{I}$ $(C,T_{\omega},PP)\xlongequal{?} 1$.

Suppose that there exists a PPT adversary $\mathcal{A}$ that can break the keyword-privacy of our scheme with the advantage at least $\epsilon'(\lambda)$,
we can construct an algorithm $\mathcal{B}$ which can use $\mathcal{A}$ to break the IBE-ANO-CPA security of the IBE scheme \cite{Boyen_2006} as follows.
Assume that $\mathcal{C}$ is the challenger in the IBE-ANO-CPA security game.\par

\textbf{Setup.} After obtaining
the public parameters $PP_{I}=(G,G_T,e,p,g,g_0,g_1)$ from $\mathcal{C}$,
$\mathcal{B}$ selects random $h_1,h_2 \stackrel{R}{\leftarrow} G$.
Let $H: x \to g_0 g_1^{x} $ be a function from $Z_p$ to $G$
and $H_1:\{0,1\}^{*} \to Z_p$ be a cryptographic hash function.
$\mathcal{B}$ generates a table $Table_{\omega}$ which includes
$(\omega,g^{\omega})$ for all keywords $\omega$ in the system.
$\mathcal{B}$ sends $(PP_{I},h_1,h_2,H,H_1,Table_{\omega})$ to $\mathcal{A}$.\par

\textbf{Phase 1.} $\mathcal{A}$ can adaptively make the following queries.\par

{\bf Key-Generation Query.} $\mathcal{A}$ adaptively makes the following queries:
\begin{enumerate}[1)]
    \item $KeyGen_{CA}$ $Query$. $\mathcal{B}$ selects $x,y \stackrel{R}{\leftarrow} Z_p$ and computes $X=g^x,Y=g^y$.
    Let $sk_{CA}=(x,y)$, $pk_{CA}=(X,Y)$.
    $\mathcal{B}$ sends the public key $pk_{CA}$ to $\mathcal{A}$.
    \item $KeyGen_{TGC}$ $Query$. After $\mathcal{B}$ is given $mpk_{I}=(\Omega,\nu_1,$ $\nu_2,\nu_3,\nu_4)$ by $\mathcal{C}$.
    Let $pk_{TGC}=mpk_{I}$.
    Then $\mathcal{B}$ sends the public key $pk_{TGC}$ to $\mathcal{A}$.
    \item $KeyGen_{TR}$ $Query$. $\mathcal{B}$ selects $x_t \stackrel{R}{\leftarrow} Z_p$
    and computes $Y_t=g^{x_t}$. $\mathcal{B}$ sends the public key $Y_t$ to $\mathcal{A}$.
    \item $KeyGen_U$ $Query$. $\mathcal{B}$ selects $x_u \stackrel{R}{\leftarrow} Z_p$ and computes $Y_u=g^{x_u}$.
    $\mathcal{B}$ sends the public key $Y_u$ to $\mathcal{A}$.
\end{enumerate}
\par

\textbf{Registration Query.} $\mathcal{A}$ adaptively submits
a $\mathcal{DU}$'s public key $Y_u$ with $\varPi _1:PoK\{(x_u):Y_u=g^{x_u}\}$ and $ID_U \in \{0,1\}^{*}$ to $\mathcal{B}$.
Then $\mathcal{B}$ verifies the proof $\varPi_1$. If it is correct,
$\mathcal{B}$ chooses $r_u \stackrel{R}{\leftarrow} Z_p$ and computes $a=g^{r_u}$, $b=a^y$, $c=a^{x}Y_u^{r_uxy}$.
$\mathcal{B}$ sends $(a,b,c)$ to $\mathcal{A}$. \par

\textbf{Trapdoor Query.} $\mathcal{B}$ selects $\hat{r_1},\hat{r_2},t_1',t_2',t_3',t_4' \stackrel{R}{\leftarrow} Z_p$.
$\mathcal{A}$ selects $u_0,u_1,u_2,r'_1,r'_2 \stackrel{R}{\leftarrow} Z_p $ and runs TTP-Module Arithmetics \cite{Jarecki_2009} protocol with $\mathcal{B}$. The results are three randomly distributed values
$x_0 =(\hat{r_1}r'_1t_1't_2'+\hat{r_2}r'_2t_3't_4')+u_0$, $x_1=-(u_3/r'_1\cdot t_0' t_2')+u_1$, $x_2=-(u_3/r'_1 \cdot t_0' t_1')+u_2$.
Notably, in the two party protocol, $\mathcal{B}$ can provide random inputs. Using rewinding techniques,
$\mathcal{B}$ can extract $u_0,u_1,u_2,r'_1,r'_2$.
Then $\mathcal{A}$ selects $r_0,r,r',u_3\stackrel{R}{\leftarrow} Z_p $ and a keyword $\omega$. $\mathcal{A}$ computes $H(\omega)' =(g_0g_1^{\omega})^{u_3}$,
$D_1=g^{\omega}Y^{r_0}_t$, $D_2=Y_uY^{r_0}_t$, $D_3=g^{r_0}$, $D_4=g^{\omega}h^{r}_1$,
$D_5=Y_uh^{r}_2$, $\tilde{\sigma}=(a^{r'},b^{r'}, c^{r'r})=(\tilde{a},\tilde{b},\hat{c})$ with a zero-knowledge proof
$\varPi _2:PoK\{(\omega,u_3,r_0,r,x_u):
H(\omega)' =(g_0g_1^{\omega})^{u_3} \wedge D_1=g^{\omega}Y^{r_0}_t
\wedge D_2=g^{x_u}Y^{r_0}_t \wedge D_3=g^{r_0}
\wedge D_4=g^{\omega}h^{r_5}_1 \wedge g^{x_u}h^{r}_2
\wedge v^{r^{-1}}_s=v_x v^{x_u}_{xy}\}$
where $v_x=e(X,\tilde{a}),v_{xy}=e(X,\tilde{b})$ and $v_s=e(g,\hat{c})$.
$\mathcal{A}$ sends $R_U=(H(\omega)',D_1,D_2,D_3,D_4,D_5,\varPi _2)$ to $\mathcal{B}$.
Then $\mathcal{B}$ stores $R_U$ on $\mathcal{BC}$. \par

$\mathcal{B}$ uses the rewinding techniques to extract $\omega$ and $u_3$,
then verifies $\varPi_2$ and $e(\tilde{a},Y)\xlongequal{?}e(g,\tilde{b})$.
If they are correct, $\mathcal{B}$ makes secret key query for keyword $\omega$ to $\mathcal{C}$.
$\mathcal{C}$ runs $Extract_{I}$ and returns the secret key $(d_0,d_1,d_2,d_3,d_4)$.
After obtaining $(d_0,d_1,d_2,d_3,d_4)$ from $\mathcal{C}$,
$\mathcal{B}$ computes $d_0'=d_0 \cdot g^{u_0}$, $d_1'=d_1^{u_3/r_1'}g^{u_1}$,
$d_2'=d_2^{u_3/r_1'}g^{u_2}$, $d_3'=d_3^{u_3/r_2'}$, $d_4'=d_4^{u_3/r_2'}$.
After that, $\mathcal{B}$ sends $(d_0',d_1',d_2',d_3',d_4')$ to $\mathcal{A}$.
Finally, $\mathcal{A}$ can get
$T_{\omega}=(d_0=d_0'\cdot g^{-u_0},d_1=(d_1' \cdot g^{-u_1})^{r_1'/u_3},d_2=(d_2' \cdot g^{-u_2})^{r_1'/u_3},d_3=d_3'^{r_2'/u_3},d_4=d_4'^{r_2'/u_3})$.
Let $T_{trapdoor}$ be an initially empty set which consists of keywords. $\mathcal{B}$ adds $\omega$ into $T_{trapdoor}$.
\par

\textbf{Challenge.} $\mathcal{A}$ submits two keywords $\omega_0,\omega_1 \notin T_{trapdoor}$ to $\mathcal{B}$.
$\mathcal{B}$ chooses a random $M \in G_T$. Then $\mathcal{B}$ sends $\omega_0$, $\omega_1$ and $M$ to
$\mathcal{C}$ as the challenge in IBE-ANO-CPA game.
$\mathcal{C}$ flips an unbiased coin with $\{0,1\}$ and obtains a bit $\mu \in \{0,1\}$.
Then $\mathcal{C}$ uses $\textsf{Encrypt}_\textsf{I}$ to computes ciphertext of $M$ with $\omega_{\mu}$ as public key.
After obtaining the ciphertext $(\hat{C},C_0,C_1,C_2,C_3,C_4)$ from $\mathcal{C}$,
$\mathcal{B}$ computes $C'=\hat{C}/M$ and sends $(C',C_0,C_1,C_2,C_3,C_4)$ to $\mathcal{A}$.

\textbf{Phase 2.} $\mathcal{A}$ continues to make the key-generation queries, registration queries and trapdoor queries,
expect the trapdoor queries for $\omega_0$ and $\omega_1$.

\textbf{Output.} $\mathcal{A}$ returns its guess $\mu'$ on $\mu$ in the game.
Then $\mathcal{B}$ outputs $\mu'$ as its guess in IBE-ANO-CPA game.
\par
Next, the advantage $\epsilon(\lambda)$ with which $\mathcal{B}$ can wins IBE-ANO-CPA game can be computed as follows.
If $\mu=\mu'$, $\mathcal{A}$ wins the above game and $\mathcal{B}$ wins IBE-ANO-CPA game.
Because $\mathcal{A}$ wins the above game with advantage $\epsilon'(\lambda)$,
the advantage with which $\mathcal{B}$ wins the IBE-ANO-CPA game is
$\epsilon(\lambda) \geq \epsilon'(\lambda)$.

\begin{theorem}Our scheme is $(t',\epsilon'(\lambda))$-keyword-blindness if the $(t,\epsilon(\lambda))$-DDH assumption holds on $(G,p,g)$,
where $\epsilon(\lambda)\geq \frac{\epsilon'(\lambda)}{2}$, $t=O(t')$.\end{theorem}
\par

\textit{Proof:} Suppose that there exists a PPT adversary $\mathcal{A}$ that can break the keyword-blindness of our scheme with the advantage at least $\epsilon'(\lambda)$, we can construct a PPT algorithm
$\mathcal{B}$ which can use $\mathcal{A}$ to break the DDH assumption as follows. $\mathcal{C}$ flips an unbiased coin with $\{0,1\}$ and obtains a bit $\mu \in \{0,1\}$.
If $\mu=1$, $\mathcal{C}$ sends $(g,g^\alpha,g^\beta,Z=g^{\alpha \beta})$ to $\mathcal{B}$; otherwise, sends $(g,g^\alpha,g^\beta,Z=R)$ to $\mathcal{B}$ where $R\in G$ is a random element in $G$.
$\mathcal{B}$ will outputs her guess $\mu'$ on $\mu$.\par

\textbf{Setup.} $\mathcal{B}$ runs $\mathcal{BG}(1^l)\to(G,G_T,e,p)$ with $e:G \times G \to G_T$.
$\mathcal{B}$ selects random $g,g_0,g_1,h_1,h_2 \stackrel{R}{\leftarrow} G$.
Let $H: x \to g_0 g_1^{x} $ be a function from $Z_p$ to $G$
and $H_1:\{0,1\}^{*} \to Z_p$ be a cryptographic hash function.
$\mathcal{CA}$ generates a table $Table_{\omega}$ which includes
$(\omega,g^{\omega})$ for all keywords $\omega$ in this system.
$\mathcal{B}$ sends $(G,G_T,e,p,g,g_0,g_1,h_1,h_2,H,H_1,Table_{\omega})$ to $\mathcal{A}$.\par

\textbf{Key-Generation Query.} $\mathcal{A}$ adaptively makes the following queries:
\begin{enumerate}[1)]
    \item $KeyGen_{CA}$ $Query$. $\mathcal{B}$ selects $x,y \stackrel{R}{\leftarrow} Z_p$ and computes $X=g^x,Y=g^y$.
    Let $sk_{CA}=(x,y)$, $pk_{CA}=(X,Y)$.
    $\mathcal{B}$ sends the public key $pk_{CA}$ to $\mathcal{A}$.
    \item $KeyGen_{TGC}$ $Query$. $\mathcal{B}$ selects $t_0, t_1, t_2, t_3, t_4 \stackrel{R}{\leftarrow} Z_p$
    and computes $\Omega=e(g,g)^{t_0 t_1 t_2}$, $\nu_1=g^{t_1}$, $\nu_2=g^{t_2}$, $\nu_3=g^{t_3}$, $\nu_4=g^{t_4}$.
    Let $pk_{TGC}=(\Omega,\nu_1,\nu_2,\nu_3,\nu_4)$, $sk_{TGC}=(t_0, t_1, t_2, t_3, t_4)$.
    $\mathcal{B}$ sends the secret-public key $(sk_{TGC},pk_{TGC})$ to $\mathcal{A}$.
    \item $KeyGen_{TR}$ $Query$. $\mathcal{B}$ lets $Y_t=g^{\alpha}$. By doing this, $\mathcal{B}$ implicitly defines secret is $\alpha$.
    $\mathcal{B}$ sends the public key $Y_t$ to $\mathcal{A}$.
    \item $KeyGen_U$ $Query$. $\mathcal{B}$ selects $x_u \stackrel{R}{\leftarrow} Z_p$ and computes $Y_u=g^{x_u}$.
    $\mathcal{B}$ sends the secret-public key pair $(x_u,Y_u)$ to $\mathcal{A}$.
\end{enumerate}\par

\textbf{Registration Query.} $\mathcal{A}$ adaptively submits a $\mathcal{DU}$'s public key $Y_u$ with $\varPi _1:PoK\{(x_u):Y_u=g^{x_u}\}$ and $ID_U \in \{0,1\}^{*}$ to $\mathcal{B}$.
Then $\mathcal{B}$ verifies the proof $\varPi_1$. If it is correct,
$\mathcal{B}$ chooses $r_u \stackrel{R}{\leftarrow} Z_p$ and computes $a=g^{r_u}$, $b=a^y$, $c=a^{x}Y_u^{r_uxy}$. $\mathcal{B}$ sends $\sigma_U=(a,b,c)$ to $\mathcal{A}$. \par

\textbf{Trapdoor Query.}  $\mathcal{B}$ selects $u_0,u_1,u_2,r'_1,r'_2 \stackrel{R}{\leftarrow} Z_p $.
Then $\mathcal{A}$ adaptively submits $\hat{r_1}$, $\hat{r_2}$ to $\mathcal{B}$ and runs TTP-Module Arithmetics \cite{Jarecki_2009} protocol with $\mathcal{B}$, the result
of which are three randomly distributed values $x_0 =(\hat{r_1}r'_1t_1t_2+\hat{r_2}r'_2t_3t_4)+u_0$, $x_1=-(u_3/r'_1\cdot t_0 t_2)+u_1$, $x_2=-(u_3/r'_1 \cdot t_0 t_1)+u_2$.
$\mathcal{B}$ selects $r_0,r,r',u_3,x_u,r_u \stackrel{R}{\leftarrow} Z_p$ and a random keyword $\omega$.
After that, $\mathcal{B}$ computes $H(\omega)' =(g_0g_1^{\omega})^{u_3}$,
$D_1=g^{\omega}Y^{r_0}_t$, $D_2=Y_uY^{r_0}_t$, $D_3=g^{r_0}$, $D_4=g^{\omega}h^{r}_1$,
$D_5=Y_uh^{r}_2$, $\tilde{\sigma}=(a^{r'},b^{r'}, c^{r'r})=(\tilde{a},\tilde{b},\hat{c})$ where $a=g^{r_u}$,$b=a^y$,$c=a^xg^{x_ur_uxy}$
and generates a zero-knowledge proof
$\varPi _2:PoK\{(\omega,u_3,r_0,r,x_u):
H(\omega)' =(g_0g_1^{\omega})^{u_3} \wedge D_1=g^{\omega}Y^{r_0}_t
\wedge D_2=g^{x_u}Y^{r_0}_t \wedge D_3=g^{r_0}
\wedge D_4=g^{\omega}h^{r}_1 \wedge D_5=g^{x_u}h^{r}_2
\wedge v^{r^{-1}}_s=v_x v^{x_u}_{xy}\}$
where $v_x=e(X,\tilde{a}),v_{xy}=e(X,\tilde{b})$
and $v_s=e(g,\hat{c})$.
Last, $\mathcal{B}$ sends  $R_U=(H(\omega)',D_1,D_2,D_3,D_4,D_5,\varPi _2)$ to $\mathcal{A}$
and $\mathcal{A}$ stores $R_U$ on $\mathcal{BC}$. \par

\textbf{Challenge.}  $\mathcal{A}$ submits two keywords $\omega_0$ and $\omega_1$.
$\mathcal{B}$ flips an unbiased coin with $\{0,1\}$ and obtains a bit $\theta  \in \{0,1\}$.
Then $\mathcal{B}$ selects $r,r',u_3,x_u,r_u \stackrel{R}{\leftarrow} Z_p$ and
computes $H(\omega_\theta)'^{*} =(g_0g_1^{\omega_\theta})^{u_3}$,
$D_1^{*}=g^{\omega_\theta} \cdot Z$, $D_2^{*}=g^{x_u} \cdot Z$, $D_3^{*}=g^{\beta}$, $D_4^{*}=g^{\omega_\theta}h_1^{r}$,
$D_5^{*}=g^{x_u}h_2^{r}$, $\tilde{\sigma}^{*}=(a^{r'},b^{r'}, c^{r'r})=(\tilde{a},\tilde{b},\hat{c})$ where $a=g^{r_u}$,$b=a^y$,$c=a^xg^{x_ur_uxy}$.
Meanwhile, $\mathcal{B}$ can simulate a zero-knowledge proof
$\varPi _2^{*}:PoK\{(\omega_\theta,u_3,\beta,r,x_u):
H(\omega_\theta)'^{*} =(g_0g_1^{\omega_\theta})^{u_3} \wedge D_1^{*}=g^{\omega_\theta}Y^{\beta}_t
\wedge D_2^{*}=g^{x_u}Y^{\beta}_t \wedge D_3^{*}=g^{\beta}
\wedge D_4^{*}=g^{\omega_\theta}h^{r_5}_1 \wedge D_5^{*}=g^{x_u}h^{r}_2
\wedge v^{r^{-1}}_s=v_x v^{x_u}_{xy}\}$
where $v_x=e(X,\tilde{a}),v_{xy}=e(X,\tilde{b})$
and $v_s=e(g,\hat{c})$.
Let $R_U^{*}=(H(\omega_\theta)'^{*},D_1^{*},D_2^{*},D_3^{*},D_4^{*},D_5^{*},\tilde{\sigma}^{*},\varPi _2^{*})$.
$\mathcal{B}$ sends $R_U^{*}$ to $\mathcal{A}$.

\textbf{Output.} $\mathcal{A}$ outputs his guess $\theta'$ on $\theta$. If $\theta=\theta'$, $\mathcal{B}$ outputs $\mu'=1$;
otherwise, $\mathcal{B}$ outputs $\mu'=0$.\par

Next, the advantage $\epsilon(\lambda)$ with which $\mathcal{B}$ can break the DDH assumption can be computed as follows.
If $\mu=1$, $Z=g^{\alpha \beta}$. Hence, $R_U^{*}$ is valid.
$\mathcal{A}$ outputs $\theta=\theta'$ with probability at least $\frac{1}{2}+\epsilon'(\lambda)$.
Since $\theta=\theta'$, $\mathcal{B}$ outputs $\mu'=1$.
We have $Pr[\mu=\mu'|\mu=1]\geq \frac{1}{2}+\epsilon'(\lambda)$.
If $\mu=0$, $Z=R$ where $R$ is a random element in $G$.
Hence, $R_U^{*}$ is invalid. $\mathcal{A}$ outputs $\theta \neq \theta'$ with probability $\frac{1}{2}$.
Since $\theta \neq \theta'$, $\mathcal{B}$ outputs $\mu'=0$. We have $Pr[\mu=\mu'|\mu=0] = \frac{1}{2}$.
The advantage with which $\mathcal{B}$ breaks the DDH assumption is
$\left\lvert \frac{1}{2}Pr[\mu=\mu'|\mu=1]-\frac{1}{2}Pr[\mu=\mu'|\mu=0] \right\rvert \geq \frac{1}{2}Pr[\mu=\mu'|\mu=1]-\frac{1}{2}Pr[\mu=\mu'|\mu=0] = \frac{1}{2} \times (\frac{1}{2}+\epsilon'(\lambda)) - \frac{1}{2} \times \frac{1}{2} =\frac{\epsilon'(\lambda)}{2}$.\par

Therefore, the advantage with which $\mathcal{B}$ can break the DDH assumption is at least $\frac{\epsilon'(\lambda)}{2}$, namely $\epsilon(\lambda)\geq \frac{\epsilon'(\lambda)}{2}$.

\begin{theorem} \label{theorem:unforgeability} Our scheme is $(t',\epsilon'(\lambda))$-unforgeability if the $(t,\epsilon(\lambda))$-LRSW
assumption holds on the bilinear group $(G,G_T,e,p)$,
where $\epsilon(\lambda)\geq \epsilon'(\lambda)$ and $t=O(t')$.
\end{theorem}\par

\textit{Proof:} Suppose that there exists a PPT adversary $\mathcal{A}$ that can break the unforgeability of our scheme with the advantage at least $\epsilon'(\lambda)$, we can construct a PPT algorithm
$\mathcal{B}$ which can use $\mathcal{A}$ to break the LRSW assumption as follows. Given a bilinear group $(G,G_T,e,p,g)$
where $g$ is a generator of $G$, public key $X=g^x,Y=g^y$ and
an oracle $O_{X,Y}(\cdot)$ to answer a query $m \in Z_p$ with a triple $(a,a^{y},a^{x+mxy})$ for a random group element $a \in G$,
$\mathcal{B}$ can output $(m,a,a^{y},a^{x+mxy})$ with restriction of $m \neq 0$ and $m$ was not queried previously. \par
\textbf{Setup.}
$\mathcal{B}$ randomly selects
$g_0,g_1,h_1,h_2 \stackrel{R}{\leftarrow} G$.
Let $H: x \to g_0 g_1^{x} $ be a function from $Z_p$ to $G$
and $H_1:\{0,1\}^{*} \to Z_p$ be a cryptographic hash function.
$\mathcal{CA}$ generates a table $Table_{\omega}$ which includes
$(\omega,g^{\omega})$ for all keywords $\omega$ in this system.
$\mathcal{B}$ sends $(G,G_T,e,p,g,g_0,g_1,h_1,h_2,H,H_1,Table_{\omega})$ to $\mathcal{A}$.\par

\textbf{Key-Generation Query.} $\mathcal{A}$ adaptively makes the following queries:
\begin{enumerate}[1)]
    \item $KeyGen_{CA}$ $Query$. $\mathcal{B}$ sends the public key $(X,Y)$ in LRSW assumption to $\mathcal{A}$.
    By doing this, $\mathcal{B}$ implicitly defines $pk_{CA}=(X,Y)$.
    \item $KeyGen_{TGC}$ $Query$. $\mathcal{B}$ selects $t_0, t_1, t_2, t_3, t_4 \stackrel{R}{\leftarrow} Z_p$
    and computes $\Omega=e(g,g)^{t_0 t_1 t_2}$, $\nu_1=g^{t_1}$, $\nu_2=g^{t_2}$, $\nu_3=g^{t_3}$, $\nu_4=g^{t_4}$.
    Let $pk_{TGC}=(\Omega,\nu_1,\nu_2,\nu_3,\nu_4)$, $sk_{TGC}=(t_0, t_1, t_2, t_3, t_4)$.
    $\mathcal{B}$ sends the public key $(sk_{TGC},pk_{TGC})$ to $\mathcal{A}$.
    \item $KeyGen_{TR}$ $Query$. $\mathcal{B}$ selects $x_t \stackrel{R}{\leftarrow} Z_p$ and computes $Y_t=g^{x_t}$.
    $\mathcal{B}$ sends the secret-public key pair $(x_t,Y_t)$ to $\mathcal{A}$.
    \item $KeyGen_U$ $Query$. $\mathcal{B}$ selects $x_u \stackrel{R}{\leftarrow} Z_p$ and computes $Y_u=g^{x_u}$.
    $\mathcal{B}$ sends the secret-public key pair $(x_u,Y_u)$ to $\mathcal{A}$.
\end{enumerate}

\textbf{Registration Query.} $\mathcal{A}$ adaptively submits a $\mathcal{DU}$'s
public key $Y_u$ with $\varPi _1:PoK\{(x_u):Y_u=g^{x_u}\}$ and $ID_U \in \{0,1\}^{*}$ to $\mathcal{B}$.
$\mathcal{B}$ verifies the proof $\varPi_1$. Using the rewinding technique, $\mathcal{B}$ can extract $x_u$.
Then $\mathcal{B}$ queries $O_{X,Y}$ for $x_u$.
After obtaining $\sigma_U=(a,a^{y},a^{x+x_uxy})$ from $O_{X,Y}$, $\mathcal{B}$ sends  $\sigma_U$ to $\mathcal{A}$.
Let $T_{pk_U}$ be a set consisting of $\mathcal{DU}$s' public keys and initially empty.
Then, $\mathcal{B}$ adds $Y_u$ into $T_{pk_U}$.\par

\textbf{Trapdoor Query.} $\mathcal{B}$ selects $u_0,u_1,u_2,r'_1,r'_2 \stackrel{R}{\leftarrow} Z_p $.
        Then $\mathcal{A}$ adaptively submits $\hat{r_1}$, $\hat{r_2}$ to $\mathcal{B}$ and
        runs TTP-Module Arithmetics \cite{Jarecki_2009} protocol with $\mathcal{B}$, the result
        of which are three randomly distributed values $x_0 =(\hat{r_1}r'_1t_1t_2+\hat{r_2}r'_2t_3t_4)+u_0$,
        $x_1=-(u_3/r'_1\cdot t_0 t_2)+u_1$, $x_2=-(u_3/r'_1 \cdot t_0 t_1)+u_2.$
        $\mathcal{B}$ selects $r_1,r,r',u_3,x_u,r_u\stackrel{R}{\leftarrow} Z_p$ and a random keyword $\omega$.
        $\mathcal{B}$ queries $O_{X,Y}$ for $x_u$ and obtains $(a, b=a^y, c=a^{x+x_uxy})$.
        Then $\mathcal{B}$ computes $H(\omega)' =(g_0g_1^{\omega})^{u_3}$, $D_1=g^{\omega}Y^{r_0}_t$, $D_2=g^{x_u}Y^{r_0}_t$, $D_3=g^{r_0}$, $D_4=g^{\omega}h_1^r$,
        $D_5=g^{x_u}h^{r}_2$, $\tilde{\sigma}=(a^{r'},b^{r'},c^{r'r})=(\tilde{a},\tilde{b},\hat{c}) $.

        After that, $\mathcal{B}$ sends $R_U=(H(\omega)',D_1,D_2,D_3,D_4,D_5,\tilde{\sigma})$
        with $\varPi _2:PoK\{(\omega,u_3,r_0,r,x_u):
        H(\omega)' =(g_0g_1^{\omega})^{u_3} \wedge D_1=g^{\omega}Y^{r_0}_t
        \wedge D_2=g^{x_u}Y^{r_0}_t \wedge D_3=g^{r_0}
        \wedge D_4=g^{\omega}h^{r}_1 \wedge D_5=g^{x_u}h^{r}_2
        \wedge v^{r^{-1}}_s=v_x v^{x_u}_{xy}\}$
        where $v_x=e(X,\tilde{a}),v_{xy}=e(X,\tilde{b})$
        and $v_s=e(g,\hat{c})$ to $\mathcal{A}$.
        Then, $\mathcal{A}$ stores $R_U$ on $\mathcal{BC}$ and $\mathcal{B}$ adds $g^{x_u}$ into $T_{pk_U}$.

        $\mathcal{A}$ verifies $\varPi_2$ and $e(\tilde{a},Y)\xlongequal{?} e(g,\tilde{b})$. If they are correct, $\mathcal{A}$ computes $d'_0=g^{x_0}$, $d'_1=g^{x_1}H(\omega)'^{-\hat{r_1}t_2}$,
        $d'_2=g^{x_2}H(\omega)'^{-\hat{r_1}t_1}$, $d'_3=H(\omega)'^{-\hat{r_2}t_4}$, $d'_4=H(\omega)'^{-\hat{r_2}t_3}$
        and sends them to $\mathcal{B}$.
        Last, $\mathcal{B}$ can get $T_{\omega}=(d_0=d_0'\cdot g^{-u_0},d_1=(d_1' \cdot g^{-u_1})^{r_1'/u_3},d_2=(d_2' \cdot g^{-u_2})^{r_1'/u_3},d_3=d_3'^{r_2'/u_3},d_4=d_4'^{r_2'/u_3})$.
        \par
\textbf{Output.} $\mathcal{A}$ outputs $R_U^{*}=(H(\omega)'^{*},D_1^{*},D_2^{*},D_3^{*},D_4^{*},D_5^{*},$ $\tilde{\sigma}^{*},\varPi_2^{*})$ and
$\varPi _2^{*}:PoK\{(\omega,u_3,r_0,r,x_u):
H(\omega)'^{*} =(g_0g_1^{\omega})^{u_3} \wedge D_1^{*}=g^{\omega}Y^{r_0}_t
\wedge D_2^{*}=g^{x_u}Y^{r_0}_t \wedge D_3^{*}=g^{r_0}
\wedge D_4^{*}=g^{\omega}h^{r}_1 \wedge D_5^{*}=g^{x_u}h^{r}_2
\wedge v^{r^{-1}}_s=v_x v^{x_u}_{xy}\}$
where $v_x=e(X,\tilde{a}^{*}),v_{xy}=e(X,\tilde{b}^{*})$
and $v_s=e(g,\hat{c}^{*})$.
$\mathcal{B}$ verifies $\varPi_2$ and $e(\tilde{a}^{*},Y)\xlongequal{?}e(g,\tilde{b}^{*})$. If they are correct,
$\mathcal{B}$ uses $sk_{TR}$ to trace and obtain $Y_u^{*}=\frac{D_2^{*}}{(D_3^{*})^{x_t}}$.
If $Y_u^{*} \in T_{pk_U}$, $\mathcal{B}$ aborts; otherwise, $\mathcal{A}$ wins the game.
$\mathcal{B}$ can use the rewinding technique to extract  $r^{-1},x_u \in Z_p$ such that $v^{r^{-1}}_s=v_x v^{x_u}_{xy}$.
Because $e(g,(\hat{c}^{*})^{r^{-1}})=e(X,\tilde{a}^{*})e(X,\tilde{b}^{*})^{x_u}$ and $e(\tilde{a}^{*},Y)=e(g,\tilde{b}^{*})$,
$(\tilde{a}^{*},\tilde{b}^{*},(\hat{c}^{*})^{r^{-1}})$ is a signature on $x_u$.
Let $\tilde{a}^{*}=g^{\alpha}$, $\tilde{b}^* = g^{\beta}$, $(\hat{c}^*)^{r^{-1}}=g^{\gamma}$.
Then we can get $e(g,g)^{\alpha y}=e(\tilde{a}^{*},Y)=e(g,\tilde{b}^{*})=e(g,g)^{\beta}$,
namely $\alpha y=\beta$. In addition, $e(g,g)^{x \alpha} e(g,g)^{x_u x \beta} =e(g,g)^{\gamma}$,
namely $x\alpha +x_u x \beta = \alpha(x+x_uxy)=\gamma$.
So $\mathcal{B}$ can use $(x_u,\tilde{a}^{*},\tilde{b}^{*},(\hat{c}^*)^{r^{-1}})$ to solve LRSW assumption.

Next, the advantage $\epsilon(\lambda)$ with which $\mathcal{B}$ can break the LRSW assumption can be computed as follows.
If $\mathcal{A}$ can forge $R_U^{*}$ and win the game with advantage $\epsilon'(\lambda)$,
the advantage with which $\mathcal{B}$ can break the LRSW assumption is $\epsilon(\lambda) \geq \epsilon'(\lambda)$.

\begin{theorem}\label{thm:light} Our scheme is $(t,\epsilon(\lambda))$-traceable
    if the LRSW assumption holds on the bilinear group $(G,G_T,e,p)$ with the advantage at most $(t_1,\epsilon_1(\lambda))$,
    the DL assumption holds on the group $G$ with the advantage at most $(t_2,\epsilon_2(\lambda))$,
    where $\epsilon(\lambda) =max\{\frac{\epsilon_1(\lambda)}{2}, \frac{\epsilon_2(\lambda)}{2}\}$.
    \end{theorem}\par
    \textit{Proof:} Suppose that there exists a PPT adversary $\mathcal{A}$ that can break
    the traceable of our scheme with the advantage at least $\epsilon(\lambda)$, we can construct a PPT algorithm
    $\mathcal{B}$ which can use $\mathcal{A}$ to break the LRSW assumption or DL assumption as follows.
    Given a bilinear group $(G,G_T,e,p,g)$  where $g$ is a generator of $G$, public key $(X=g^x,Y=g^y)$ and
    an oracle $O_{X,Y}(\cdot)$ to answer a query $m \in Z_p$ with a triple $(a,a^{y},a^{x+mxy})$ for a random group element $a \in G$,
    $\mathcal{B}$ will output $(m,a,a^{y},a^{x+mxy})$ with restriction of $m \neq 0$ and $m$ was not queried previously.

    \textbf{Setup.}
    $\mathcal{B}$ selects random
    $g_0,g_1,h_1,h_2 \stackrel{R}{\leftarrow} G$. Let $H: x \to g_0 g_1^{x} $ be a function from $Z_p$ to $G$
    and $H_1:\{0,1\}^{*} \to Z_p$ be a cryptographic hash function.
    $\mathcal{B}$ generates a table $Table_{\omega}$ which includes
    $(\omega,g^{\omega})$ for all keywords $\omega$ in this system.
    $\mathcal{B}$ sends $(G,G_T,e,p,g,g_0,g_1,h_1,h_2,H,H_1,Table_{\omega})$ to $\mathcal{A}$.\par

    \textbf{Key-Generation Query.} $\mathcal{A}$ adaptively makes the following queries:
   \begin{enumerate}[1)]
        \item $KeyGen_{CA}$ $Query$. $\mathcal{B}$ sends the public key $(X,Y)$ in LRSW assumption to $\mathcal{A}$.
        By doing this, $\mathcal{B}$ implicitly defines $pk_{CA}=(X,Y)$.
        \vspace{-0.2cm}
        \item $KeyGen_{TGC}$ $Query$. $\mathcal{B}$ selects $t_0, t_1, t_2, t_3, t_4 \stackrel{R}{\leftarrow} Z_p$
        and computes $\Omega=e(g,g)^{t_0 t_1 t_2}$, $\nu_1=g^{t_1}$, $\nu_2=g^{t_2}$, $\nu_3=g^{t_3}$, $\nu_4=g^{t_4}$.
        Let $pk_{TGC}=(\Omega,\nu_1,\nu_2,\nu_3,\nu_4)$, $sk_{TGC}=(t_0, t_1, t_2, t_3, t_4)$.
        $\mathcal{B}$ sends the secret-public key pair $(sk_{TGC},pk_{TGC})$ to $\mathcal{A}$.
          \vspace{-0.2cm}
        \item $KeyGen_{TR}$ $Query$. $\mathcal{B}$ selects $x_t \stackrel{R}{\leftarrow} Z_p$ and computes $Y_t=g^{x_t}$.
        $\mathcal{B}$ sends the secret-public key pair $(x_t,Y_t)$ to $\mathcal{A}$.
          \vspace{-0.2cm}
        \item $KeyGen_U$ $Query$. $\mathcal{B}$ selects $x_u \stackrel{R}{\leftarrow} Z_p$ and computes $Y_u=g^{x_u}$.
        $\mathcal{B}$ sends the secret-public key pair $(x_u,Y_u)$ to $\mathcal{A}$.
    \end{enumerate}

    \textbf{Registration Query.} $\mathcal{A}$ adaptively submits a $\mathcal{DU}$'s
    public key $Y_u$ with $\varPi _1:PoK\{(x_u):Y_u=g^{x_u}\}$ and $ID_U \in \{0,1\}^{*}$ to $\mathcal{B}$.
    $\mathcal{B}$ verifies the proof $\varPi_1$. Using the rewinding technique, $\mathcal{B}$ can extract $x_u$.
    Then $\mathcal{B}$ queries $O_{X,Y}$ for $x_u$.
    After obtaining $\sigma_U=(a,a^{x_uy},a^{x+x_uxy})$ from $O_{X,Y}$, $\mathcal{B}$ sends $\sigma_U$ to $\mathcal{A}$.
    Let $T_{pk_{U1}}$ be a set consisting of $\mathcal{DU}$s' public keys and initially empty.
    Then, $\mathcal{B}$ adds $Y_u$ into $T_{pk_{U1}}$.\par

    \textbf{Trapdoor Query.} $\mathcal{B}$ selects $u_0,u_1,u_2,r'_1,r'_2 \stackrel{R}{\leftarrow} Z_p $.
        Then $\mathcal{A}$ adaptively submits $\hat{r_1}$, $\hat{r_2}$ to $\mathcal{B}$ and
        runs TTP-Module Arithmetics \cite{Jarecki_2009} protocol with $\mathcal{B}$, the result
        of which are three randomly distributed values $x_0 =(\hat{r_1}r'_1t_1t_2+\hat{r_2}r'_2t_3t_4)+u_0$,
        $x_1=-(u_3/r'_1\cdot t_0 t_2)+u_1$, $x_2=-(u_3/r'_1 \cdot t_0 t_1)+u_2.$
        $\mathcal{B}$ selects $r_0,r,r',u_3,x_u,r_u\stackrel{R}{\leftarrow} Z_p$ and a random keyword $\omega$.
        $\mathcal{B}$ queries $O_{X,Y}$ for $x_u$ and obtains $\sigma_U=(a, b=a^y, c=a^{x+x_uxy})$.
        Then $\mathcal{B}$ computes $H(\omega)' =(g_0g_1^{\omega})^{u_3}$, $D_1=g^{\omega}Y^{r_0}_t$, $D_2=g^{x_u}Y^{r_0}_t$, $D_3=g^{r_0}$, $D_4=g^{\omega}h_1^r$,
        $D_5=g^{x_u}h^{r}_2$, $\tilde{\sigma}=(a^{r'},b^{r'},c^{r'r})=(\tilde{a},\tilde{b},\hat{c}) $.

        After that, $\mathcal{B}$ sends $(H(\omega)',D_1,D_2,D_3,D_4,D_5,\tilde{\sigma})$ with $\varPi _2:PoK\{(\omega,u_3,r_0,r,x_u):
        H(\omega)' =(g_0g_1^{\omega})^{u_3} \wedge D_1=g^{\omega}Y^{r_0}_t
        \wedge D_2=g^{x_u}Y^{r_0}_t \wedge D_3=g^{r_0}
        \wedge D_4=g^{\omega}h^{r}_1 \wedge D_5=g^{x_u}h^{r}_2
        \wedge v^{r^{-1}}_s=v_x v^{x_u}_{xy}\}$
        where $v_x=e(X,\tilde{a}),v_{xy}=e(X,\tilde{b})$
        and $v_s=e(g,\hat{c})$ to $\mathcal{A}$.
        Let $T_{pk_{U2}}$ be a set consisting of public key selected by $\mathcal{B}$ to answer $\mathcal{A}$ and initially empty.
        Then, $\mathcal{A}$ stores $R_U$ on $\mathcal{BC}$ and $\mathcal{B}$ adds $g^{x_u}$ into $T_{pk_{U2}}$.

        $\mathcal{A}$ verifies $\varPi_2$ and $e(\tilde{a},Y)\xlongequal{?} e(g,\tilde{b})$. If they are correct, $\mathcal{A}$ computes $d'_0=g^{x_0}$, $d'_1=g^{x_1}H(\omega)'^{-\hat{r_1}t_2}$,
        $d'_2=g^{x_2}H(\omega)'^{-\hat{r_1}t_1}$, $d'_3=H(\omega)'^{-\hat{r_2}t_4}$, $d'_4=H(\omega)'^{-\hat{r_2}t_3}$
        and sends them to $\mathcal{B}$.
        Last, $\mathcal{B}$ can get $T_{\omega}=(d_0=d_0'\cdot g^{-u_0},d_1=(d_1' \cdot g^{-u_1})^{r_1'/u_3},d_2=(d_2' \cdot g^{-u_2})^{r_1'/u_3},d_3=d_3'^{r_2'/u_3},d_4=d_4'^{r_2'/u_3})$.
        \par

    \textbf{Output.} $\mathcal{A}$ outputs $R_U^{*}=(H(\omega)'^{*},D_1^{*},D_2^{*},D_3^{*},D_4^{*},D_5^{*},$ $\tilde{\sigma}^{*},\varPi_2^{*})$ and
    $\varPi _2^{*}:PoK\{(\omega,u_3,r_0,r,x_u^{*}):
    H(\omega)'^{*} =(g_0g_1^{\omega})^{u_3} \wedge D_1^{*}=g^{\omega}Y^{r_0}_t
    \wedge D_2^{*}=g^{x_u^{*}}Y^{r_0}_t \wedge D_3^{*}=g^{r_0}
    \wedge D_4^{*}=g^{\omega}h^{r}_1 \wedge D_5^{*}=g^{x_u^{*}}h^{r}_2
    \wedge v^{r^{-1}}_s=v_x v^{x_u^{*}}_{xy}\}$
    where $v_x=e(X,\tilde{a}^{*}),v_{xy}=e(X,\tilde{b}^{*})$
    and $v_s=e(g,\hat{c}^{*})$.
    $\mathcal{B}$ verifies $\varPi_2$ and $e(\tilde{a}^{*},Y)\xlongequal{?}e(g,\tilde{b}^{*})$. If they are correct,
    $\mathcal{B}$ uses $sk_{TR}$ to trace $R_U$ and obtain $Y_u^{*}=\frac{D_2^{*}}{(D_3^{*})^{x_t}}$.
    If $Y_u^{*} \in T_{pk_{U1}}$, $\mathcal{B}$ aborts; otherwise,
    the following two types forgers are considered. In Type-1, forger outputs $R_U^{*}$ which can be traced to $Y_u^{*} \notin T_{pk_{U2}}$,
    namely $Y_u^{*}$ is not $\mathcal{DU}$s' public keys.
    In Type-2,  forger outputs $R_U^{*}$ which can be traced to $Y_u^{*} \in T_{pk_{U2}}$, namely $x_u^{*}$ is not known by $\mathcal{A}$.
    \begin{itemize}
        \item \textit{Type-1.} $Y_u^{*} \notin T_{pk_{U2}}$.
        If $\mathcal{A}$ can generate $H(\omega)'^{*},D_1^{*},D_2^{*},$ $D_3^{*},D_4^{*},D_5^{*},\tilde{\sigma}^{*}=(\tilde{a}^{*},\tilde{b}^{*},\hat{c}^{*})$
        and a zero-knowledge proof
        $\varPi _2^{*}:PoK\{(\omega,u_3,r_0,r,x_u^{*}):
        H(\omega)'^{*} =(g_0g_1^{\omega})^{u_3} \wedge D_1^{*}=g^{\omega}Y^{r_0}_t
        \wedge D_2^{*}=g^{x_u^{*}}Y^{r_0}_t \wedge D_3^{*}=g^{r_0}
        \wedge D_4^{*}=g^{\omega}h^{r}_1 \wedge D_5^{*}=g^{x_u^{*}}h^{r}_2
        \wedge v^{r^{-1}}_s=v_x v^{x_u^{*}}_{xy}\}$
        where $v_x=e(X,\tilde{a}^{*}),v_{xy}=e(X,\tilde{b}^{*})$
        and $v_s=e(g,\hat{c}^{*})$,
    $\mathcal{B}$ can use $\mathcal{A}$ to break the LRSW assumption by using the technique in the proof of Theorem \ref{theorem:unforgeability}.

        \item \textit{Type-2.} $Y_u^{*} \in T_{pk_{U2}}$.
        If $\mathcal{A}$ can generate $H(\omega)'^{*},D_1^{*},D_2^{*},$ $D_3^{*},D_4^{*},D_5^{*},\tilde{\sigma}^{*}=(\tilde{a}^{*},\tilde{b}^{*},\hat{c}^{*})$ and zero-knowledge a proof
        $\varPi _2^{*}:PoK\{(\omega,u_3,r_0,r,x_u^{*}):
        H(\omega)'^{*} =(g_0g_1^{\omega})^{u_3} \wedge D_1^{*}=g^{\omega}Y^{r_0}_t
        \wedge D_2^{*}=g^{x_u^{*}}Y^{r_0}_t \wedge D_3^{*}=g^{r_0}
        \wedge D_4^{*}=g^{\omega}h^{r}_1 \wedge D_5^{*}=g^{x_u^{*}}h^{r}_2
        \wedge v^{r^{-1}}_s=v_x v^{x_u^{*}}_{xy}\}$
        where $v_x=e(X,\tilde{a}^{*}),v_{xy}=e(X,\tilde{b}^{*})$
        and $v_s=e(g,\hat{c}^{*})$,
    $\mathcal{B}$ can use the rewinding technique to extract $x_u^{*},r_0 \in Z_p$ such that $D_2^{*}=g^{x_u^{*}}Y^{r_0}_t$,
    namely given $(g,Y_u^{*})$, $\mathcal{B}$ can output $x_u^{*}$ such that $Y_u^{*}=g^{x_u^{*}}$.
    Hence $\mathcal{B}$ can use $\mathcal{A}$ to break the DL assumption.
    \end{itemize}

    Next, the advantage with which $\mathcal{B}$ can break the LRSW assumption or DL assumption can be computed as follows.
    By the proof of Theorem \ref{theorem:unforgeability}, $\mathcal{A}$ can win the game
    with advantage $\epsilon_1(\lambda)$ in the situation of \textit{Type-1}. In the situation of \textit{Type-2},
    $\mathcal{A}$ can win the game with advantage $\epsilon_2(\lambda)$.
    Because both \textit{Type-1} and \textit{Type-2} occur with probability $\frac{1}{2}$,
    $\mathcal{B}$ can break the LRSW assumption with the advantage $\frac{\epsilon_1(\lambda)}{2}$
    or break the DL assumption with advantage $= \frac{\epsilon_2(\lambda)}{2}$. Therefore,
    $\epsilon(\lambda) =max\{\frac{\epsilon_1(\lambda)}{2},\frac{\epsilon_2(\lambda)}{2}\}$.

\section{Performance Analysis}
\label{sec:sec7}
\begin{figure*}[!h]
    \centering
        \subfloat[\textsf{Setup}]{\label{figure:test_setup}\includegraphics[width = 0.33\textwidth]{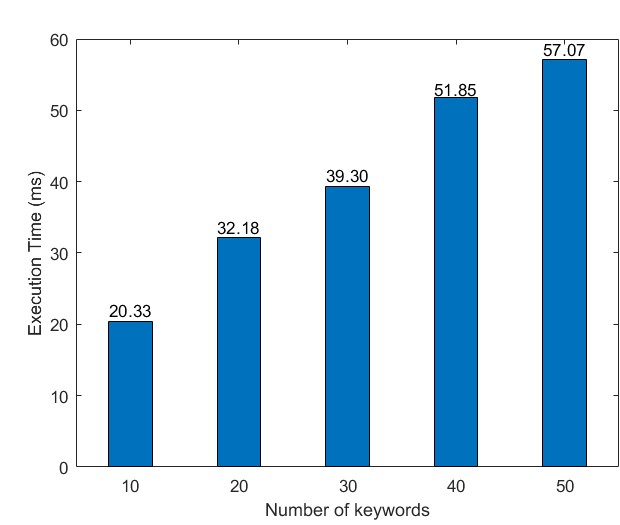}}
        \subfloat[\textsf{KeyGen}]{\label{figure:test_keygen}\includegraphics[width = 0.33\textwidth]{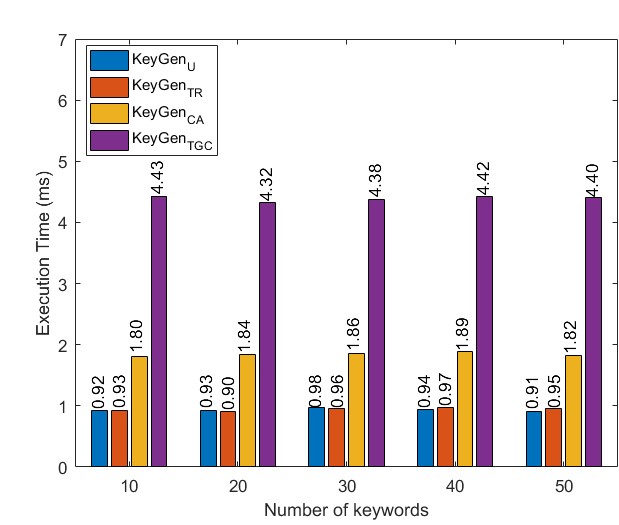}}
        \subfloat[\textsf{Reg}]{\label{figure:test_reg}\includegraphics[width = 0.33\textwidth]{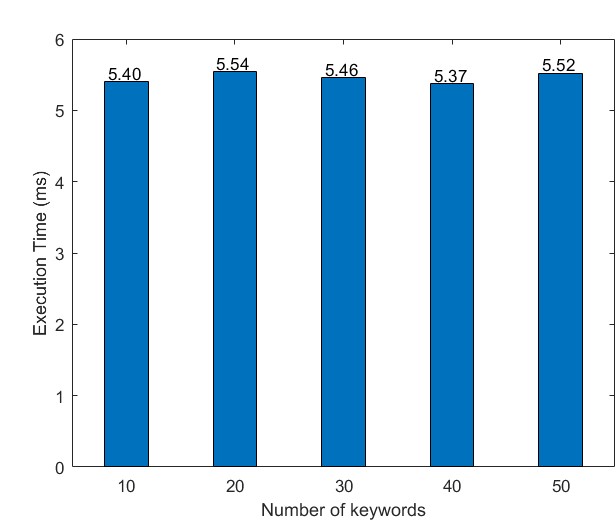}}

        \subfloat[\textsf{Trapdoor}]{\label{figure:test_trapdoor}\includegraphics[width = 0.33\textwidth]{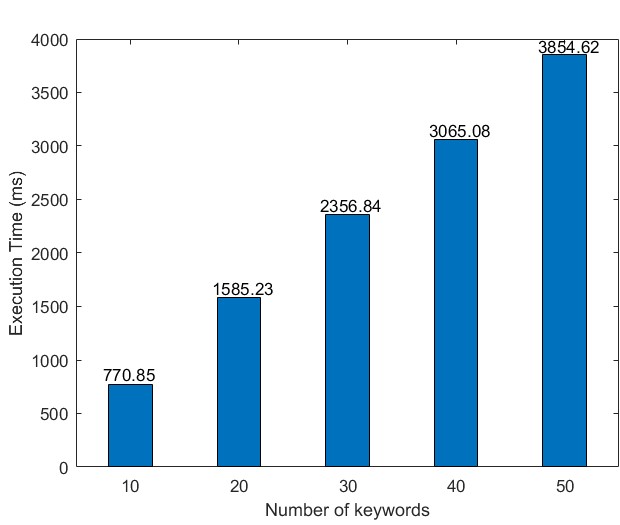}}
        \subfloat[\textsf{PEKS}]{\label{figure:test_peks}\includegraphics[width = 0.33\textwidth]{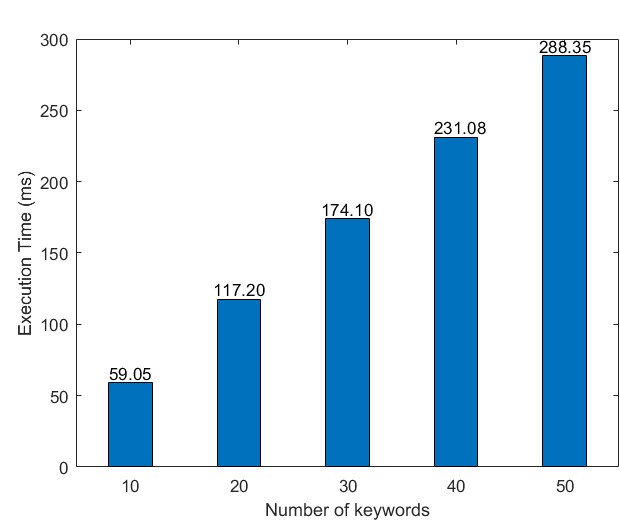}}
        \subfloat[\textsf{Test}]{\label{figure:test_test}\includegraphics[width = 0.33\textwidth]{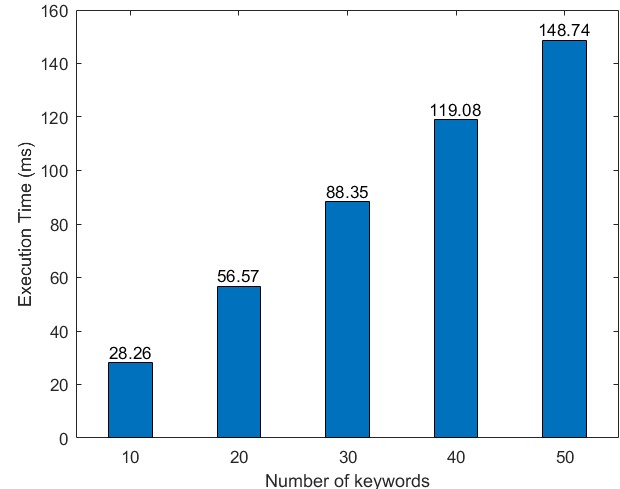}}

        \subfloat[\textsf{Record-Validation}]{\label{figure:test_valid}\includegraphics[width = 0.33\textwidth]{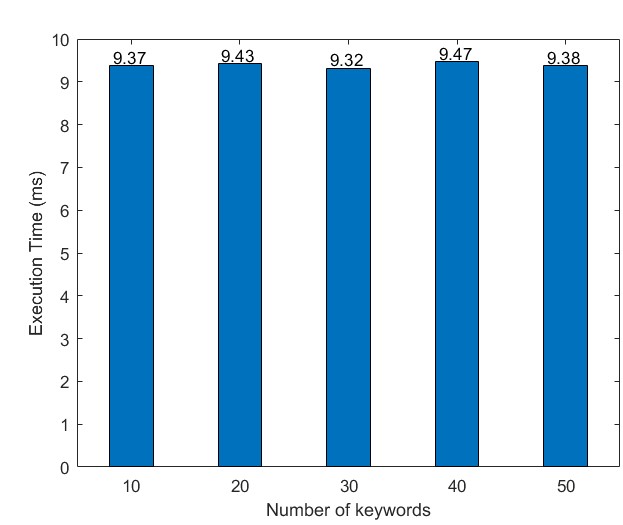}}
        \subfloat[\textsf{Trace}]{\label{figure:test_trace}\includegraphics[width = 0.33\textwidth]{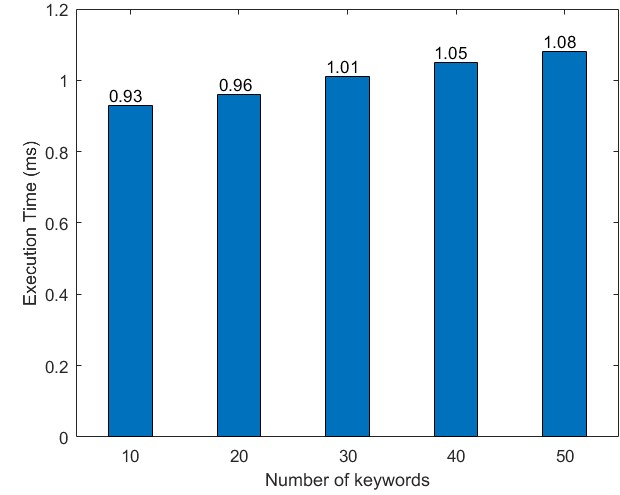}}
        \subfloat[\textsf{Gas cost}]{\label{figure:gas_cost}\includegraphics[width = 0.33\textwidth]{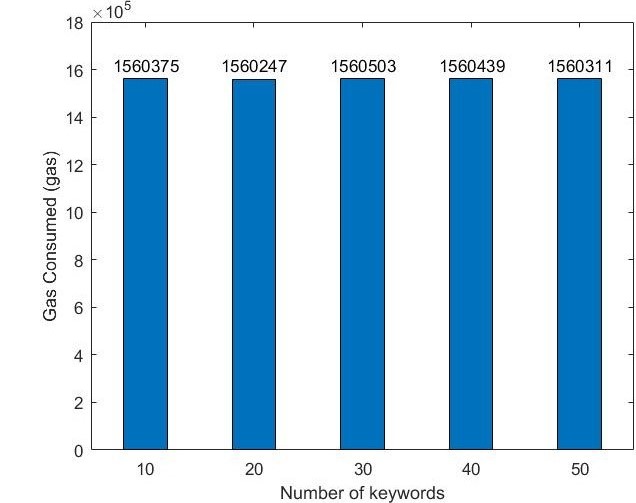}}

        \caption{The computation time of algorithms and gas cost in our BP3KSEST scheme}\label{figure:test_time}
\end{figure*}
In this section, we evaluate the performance of our scheme. The experiments are performed
on a HP 15-CX0075\\-TX laptop with an Intel i5-8300H CPU, 16G RAM and 128G SSD running
Ubuntu 18.04.6 LTS operating system in Vmware 16.2.3.
In the experiments, we used C++ language based on PBC library \cite{pbc_lib} and Openssl \cite{openssl_lib}, which  provides the
other cryptographic primitives required by our scheme.
Moreover, we apply
Type A pairing which is constructed on curve $E:y^2=x^3+x$ over the field $F_q$,
where $q$ is a large prime number and $q=3$ mod $4$.
Note that the group order of $G$ is 160 bits and the length of a group element is 512 bits.
For the hash function $H_1:\{0,1\}^{*} \to Z_p$ which is used in zero-knowledge proof, we selected SHA-256.
Besides, we used FISCO BCOS to build a local blockchain environment with 10 initial nodes to simulate our scheme.
The smart contract is written in Solidity.
The TTP-Modulo-Arithmetics protocol \cite{Jarecki_2009} was implemented
as same as \cite{Camenisch_2005} using the Paillier homomorphic encryption scheme \cite{Paillier_1999}.
The details of the proofs $\varPi_1$,$\varPi_2$ are referred to \cite{camenisch1997proof} \cite{camenisch2003practical}.
\subsection{Timing}

Suppose $n$ is the number of keywords in our scheme.
In our implementation, we executed each algorithm in the following five cases:
$(1)$ $n=10$; $(2)$ $n=20$; $(3)$ $n=30$; $(4)$ $n=40$; $(5)$ $n=50$.

In Fig. \ref{figure:test_time}, we show the computation time of the \textsf{Setup}, \textsf{KeyGen}, \textsf{Reg}, \textsf{Trapdoor}, \textsf{PEKS}, \textsf{Test},
\textsf{Record-Validation} and \textsf{Trace} algorithms in our BP3KSEST scheme. Moreover, the gas cost of storing a trapdoor query record on the blockchain is shown in Fig. \ref{figure:test_time} too.

As shown in Fig. \subref*{figure:test_setup}, the computation time of the \textsf{Setup} algorithm is 20.33 ms, 32.18 ms, 39.30 ms, 51.85 ms and 57.07 ms in the five cases, respectively.
Since a keyword table needs to be generated,
the computation time of \textsf{Setup} algorithm increases with the growth of the number of keywords.
The computation cost of the \textsf{Setup} algorithm is linear with the number $n$.

The computation time of the \textsf{KeyGen} algorithm is given in
Fig. \subref*{figure:test_keygen}, including $\textsf{KeyGen}_\textsf{U}$, $\textsf{KeyGen}_\textsf{TR}$,
$\textsf{KeyGen}_\textsf{CA}$ and
$\textsf{KeyGen}_\textsf{TGC}$ algorithms.
As described in Fig. \subref*{figure:test_keygen}, the computation costs of these algorithms are constant.

Fig. \subref*{figure:test_reg} shows the computation time of
the \textsf{Reg} algorithm for $n$ keywords.
Each data user interacts with center authority to obtain his credential by executing the \textsf{Reg} algorithm.
It takes 5.40 ms, 5.54 ms, 5.46 ms, 5.37 ms and 5.52 ms to generate a credential
for a data user in the five cases, respectively.
The computation cost of the \textsf{Reg} algorithm is constant.

Fig. \subref*{figure:test_trapdoor} shows the computation time of
the \textsf{Trapdoor} algorithm for $n$ keywords.
It takes 770.85 ms, 1585.23 ms, 2356.84 ms, 3065.08 ms and 3854.62 ms to generate
trapdoors in the five cases, respectively.
As shown in Fig. \subref*{figure:test_trapdoor}, the computation time of the \textsf{Trapdoor} algorithm
is linear with the number $n$.
Because secure two-party computation  and zero-knowledge proof are
used to achieve privacy-preserving and traceability, the computation cost of the \textsf{Trapdoor} algorithm is relatively heavy.

As shown in Fig. \subref*{figure:test_peks}, to generate encrypted keywords,
the \textsf{PEKS} algorithm takes 59.05 ms, 117.20 ms, 174.10 ms,
231.08 ms and 288.35 ms in the five cases, respectively.
The computation cost of the \textsf{PEKS} algorithm is linear with the number $n$.

In Fig. \subref*{figure:test_test}, given a trapdoor, the \textsf{Test} algorithm is executed over $n$ encrypted keywords.
The executing time of the \textsf{Test} algorithm is 28.26 ms, 56.57 ms, 88.35 ms, 119.08 ms and 148.74 ms
in the five cases, respectively.
The computation cost of the \textsf{Test} algorithm is linear with the number $n$.

As shown in Fig. \subref*{figure:test_valid}, to verify a trapdoor query record,
the \textsf{Record-Validation} algorithm takes 9.37 ms, 9.43 ms, 9.32 ms,
9.47 ms and 9.38 ms in the five cases, respectively.
The computation cost of the \textsf{Record-Validation} algorithm is constant.

Fig. \subref*{figure:test_trace} shows the computation time of
the \textsf{Trace} algorithm for $n$ keywords.
To trace a data user's identity and the keyword which he searched, 
the \textsf{Trace} algorithm  takes 0.93 ms, 0.96 ms, 1.01 ms, 1.05 ms and 1.08 ms in the five cases, respectively.
The computation cost of the \textsf{Trace} algorithm is linear with the number $n$.
The above experimental results indicate that our scheme is feasible and practical.

As shown in Fig .\subref*{figure:gas_cost}, to store a trapdoor query record on the blockchain,
the gas cost is 1560.375 $\times 10^3$ gas, 1560.274 $\times 10^3$ gas, 1560.503 $\times 10^3$ gas,
1560.439 $\times 10^3$ gas, 1560.311 $\times 10^3$ gas in the five cases, respectively.
\section{Conclusion}
\label{sec:sec8}
In this paper, a BP3KSEST
scheme was proposed to protect the privacy of both identities and keywords.
To be specific, a data user can get trapdoors from trapdoor generation center without revealing his identity and the keyword.
When trace is required,
the tracer can de-anonymise the data users and trace the keywords which they searched.
Moreover, the trapdoor query records are unforgeable and immutable in our scheme.
Furthermore, the construction, security model and security proof of our scheme are formally treated.
Finally, the experimental analysis
of our scheme was presented. Our future research will keep the properties of this work and realize
the traceability of data owners' identity and encrypted keywords they upload.

\section*{Acknowledgement}
This work was supported by the National Natural Science Foundation of China (Grant No. 62372103,
61972190, 62032005, 62002058), the Natural Science Foundation of Jiangsu Province (Grant No. BK20231149),
the Fundamental Research Funds for the Central Universities (Grant No. 2242021R40011) and the Start-up
Research Fund of Southeast University (Grant No. RF1028623300, RF1028623200).

\printcredits
\bibliographystyle{elsarticle-num}

\bibliography{paper}

\vspace{-0.3cm}
\appendix
\section{The Details of $\varPi_1$}\label{section:append1}
An instantiation of the proof $\varPi_1$ is given as follows. $\mathcal{DU}$ selects $x_u' \stackrel{R}{\leftarrow} Z_p$, and computes $Y_u'=g^{x_u'}$, $c=H_1(Y_u||Y_u')$ and
$\hat{x_u}=x_u'-c\cdot x_u$. $\mathcal{DU}$ sends $(Y_u,Y_u')$ and $(c,\hat{x_u})$ to $\mathcal{CA}$.\par
After receiving $(Y_u,Y_u',c,\hat{x_u})$. $\mathcal{CA}$ verifies $c \stackrel{?}{=} H_1(Y_u||$ $Y_u'),$ $Y_u' \stackrel{?}{=} g^{\hat{x_u}} \cdot Y_u^{c}$.\par
\vspace{-0.3cm}
\section{The Details of $\varPi_2$}\label{section:append2}
An instantiation of the proof $\varPi_2$ is given as follows. $\mathcal{DU}$ selects $\omega',u_3',r_0',r',x_u'\stackrel{R}{\leftarrow} Z_p$,
and computes $H(\omega)''=g_0^{u_3'}g_1^{\omega'u_3'}$,$D_1'=g^{\omega'}Y^{r_0'}_t$, $D_2'=g^{x_u'}Y^{r_0'}_t$, $D_3'=g^{r_0'}$,
$D_4'=g^{\omega'}h_1^{r'}$, $D_5'=g^{x_u'}h^{r'}_2$, $v_x'=v_s^{k'}v_{xy}^{-x_u'}$, $\hat{t}=\omega' u_3'-c\cdot (\omega u_3)$,
$\hat{\omega}=\omega'-c\cdot \omega$, $\hat{u_3}=u_3'-c\cdot u_3$, $\hat{r_0}=r_0'-c\cdot r_0$,
$\hat{r}=r'-c\cdot r$, $\hat{x_u}=x_u'-c\cdot x_u$,
$\hat{k}=k'-c\cdot r^{-1}$ and $c=H_1(H(\omega)'||H(\omega)''||D_1||D_1'||D_2||D_2'||D_3||D_3'||D_4||D_4'||\\D_5||D_5'||v_x||v_x')$.
Then $\mathcal{DU}$ sends $(H(\omega)'',D_1',D_2',D_3',D_4',$ $D_5',$ $v_x')$ and $(\hat{t},\hat{\omega},$
$\hat{u_3},\hat{r_0},\hat{r},\hat{x_u},\hat{k})$ to $\mathcal{TGC}$.\par

After receiving $(H(\omega)'',D_1',D_2',D_3',D_4',D_5',v_x')$ and $(c,\hat{t},\hat{\omega},\hat{u_3},\hat{r_0}, \hat{r},\hat{x_u},\hat{k})$,
$\mathcal{TGC}$ verifies
$c \stackrel{?}{=} H_1(H(\omega)'|| H(\omega)''$ $||D_1||D_1'||D_2||D_2'||D_3||D_3'||D_4||D_4'||D_5||D_5'||v_x||v_x'),\\H(\omega)''\stackrel{?}{=} g_0^{\hat{u_3}} g_1^{\hat{t}}H(\omega)'^{c}$,
$D_1' \stackrel{?}{=} g^{\hat{\omega}} Y_t^{\hat{r_0}}D_1^{c}$, $D_2' \stackrel{?}{=} g^{\hat{x_u}}Y_t^{\hat{r_0}}D_2^{c}$,
$D_3' \stackrel{?}{=} g^{\hat{r_0}}D_3^{c}$, $D_4' \stackrel{?}{=} g^{\hat{\omega}}h_1^{\hat{r}}D_4^{c}$,
$D_5' \stackrel{?}{=} g^{\hat{x_u}}h_2^{\hat{r}}D_5^{c}$,
$v_x' \stackrel{?}{=}  v_s^{\hat{k}}v_{xy}^{-\hat{x_u}}v_x^c$.

\end{document}